\documentclass[a4paper, amsfonts, amssymb, amsmath, reprint, showkeys, nofootinbib, twoside, iop]{revtex4-1}

\usepackage[utf8]{inputenc} 
\usepackage[english]{babel} 

\usepackage{hyperref} 
\usepackage{url} 
\hypersetup{            
  colorlinks   = true,  
  urlcolor     = blue,  
  linkcolor    = black, 
  citecolor   = blue    
}

\usepackage{amsmath}    
\usepackage{amsfonts}   
\usepackage{amssymb}    
\usepackage{mathtools}  
\usepackage{textcomp} 
\usepackage{mathrsfs} 
\usepackage{cancel} 
\usepackage{xcolor}

\usepackage{cleveref} 

\usepackage{booktabs} 
\usepackage[caption=false]{subfig}     
\usepackage{float} 

\usepackage{lastpage} 

\newcommand{\ee}{\mathrm{e}}
\newcommand{\sigmaVec}{\boldsymbol{\sigma}}   

\begin{document}

\graphicspath{{figures/}}

\author{Frederik L. Durhuus}
\affiliation{CAMD, Department of Physics, Technical University of Denmark, 2800 Kgs.\ Lyngby, Denmark}
\author{Thorbj\o rn Skovhus}
\affiliation{CAMD, Department of Physics, Technical University of Denmark, 2800 Kgs.\ Lyngby, Denmark}
\author{Thomas Olsen}
\email{tolsen@fysik.dtu.dk}
\affiliation{CAMD, Department of Physics, Technical University of Denmark, 2800 Kgs.\ Lyngby, Denmark}

\title{Plane wave implementation of the magnetic force theorem for magnetic exchange constants: Application to bulk Fe, Co and Ni}

\date{\today}

\begin{abstract}
We present a plane wave implementation of the magnetic force theorem, which provides a first principles framework for extracting exchange constants parameterizing a classical Heisenberg model description of magnetic materials. It is shown that the full microscopic exchange tensor may be expressed in terms of the static Kohn-Sham susceptibility tensor and the exchange-correlation magnetic field. This formulation allows one to define arbitrary magnetic sites localized to predefined spatial regions, hence rendering the problem of finding Heisenberg parameters independent of any orbital decomposition of the problem. The susceptibility is calculated in a plane wave basis, which allows for systematic convergence with respect to unoccupied bands and spatial representation. We then apply the method to the well-studied problem of calculating adiabatic spin wave spectra for bulk Fe, Co and Ni, finding good agreement with previous calculations. In particular, we utilize the freedom of defining magnetic sites to show that the calculated Heisenberg parameters are robust towards changes in the definition of magnetic sites. This demonstrates that the magnetic sites can be regarded as well defined and thus asserts the relevance of the Heisenberg model description despite the itinerant nature of the magnetic state.
\end{abstract}

\maketitle

\pagenumbering{arabic}

\section{Introduction}
Spin waves constitute fundamental excitations to the magnetic order of crystalline magnetic materials and a proper understanding of spin waves in solids is crucial for predicting a wide range of basic magnetic properties. For example, the thermodynamical properties of magnetic materials are largely governed by thermal spin wave excitations and the critical temperature for magnetic order can typically be accurately estimated if the spin wave spectrum is known. For two-dimensional magnets in particular, it is well known that Weiss mean field theory is inapplicable and that the persistence of magnetic order is directly
linked to the presence of a gap in the spin wave spectrum \cite{2D_magnetism_exp17,2D_itinerant_magnetism_exp18, Torelli2018Critical_temperatures_in_2D}. In addition, it is generally believed that spin fluctuations play a central role in certain classes of high temperature superconductors and that a detailed understanding of the spin dynamics in these materials is a highly desired piece in the puzzle of unconventional pairing mechanisms \cite{Scalapino2012,romer2020pairing,Tutueau2020magnetic_stripes_INS_experiment}. In terms of technological relevance, the field of magnonics comprises the most direct application, where the spin waves act as basic information carriers for low energy information processing \cite{Kruglyak2010magnonics_review} and bottom-up design of optimized materials relies heavily on a detailed understanding of the spin wave dispersion. Moreover, the study of magnetic excitations provide crucial information about damping mechanisms in magnetic moment dynamics \cite{Guimaraes2019Comparative_study_of_Gilbert_damping_calculations}, which are essential for optimizing power consumption and read/write speed in spintronics \cite{Bhatti2017SpintronicsRAM}.
Finally, computing magnetic excitations enables the {\it ab initio} simulation of inelastic neutron scattering spectra \cite{Etz2015ASD_and_surface_magnons}, which may accelerate material science at current and upcoming experimental facilities \cite{Andersen2020ESS_instrument_suite}. 

Theoretical predictions of the spin wave dispersion in a given magnetic material can be obtained directly from first principles using either time-dependent density functional theory \cite{runge-gross, VanSchilfgaarde1999,Pajda_2001_ab_init_exchange,Buczek2011,Singh2019,Rousseau2012,Cao2018,Skovhus2021Published_version,Skovhus2022}, many-body perturbation theory (MBPT) \cite{Karlsson1999,Kotani_2008,Sasoglu2010,Muller2016,Okumura2019,Friedrich2020,Olsen2021} or dynamical mean field theory \cite{Kotliar2006DMFT,Liechtenstein2001FiniteTemperatureMagnetismDMFT}. These methods provide direct access to the magnetic susceptibility and thus to the structure factors and magnon lifetimes as well as the spin wave dispersion. Each of the methods are, however, rather demanding in terms of computational power and typically requires careful convergence with respect to various parameters entering the calculations. As such, direct first principles computations are typically only applicable to rather simple materials containing only a small number of atoms per unit cell. If one is interested in performing a high-throughput screening of a large number of magnetic materials for particular spin wave properties, a computationally simpler and cheaper approach is paramount.

One such alternative is to fit the parameters of a classical Heisenberg model to total energy calculations from density functional theory (DFT). Given the Heisenberg parameters in reciprocal space, the spectrum of non-interacting spin waves can readily be computed from linear spin wave theory \cite{Yosida1996, Toth2015LinearSpinWaveTheory}. Spin wave interactions may subsequently be included perturbatively or by means of mean-field theory \cite{Zhitomirsky2009Spin_waves_in_triangular_AFM}. 
Heisenberg models do not include the effect of Stoner excitations (single particle spin-flip excitations) and thus cannot capture the finite spin wave lifetimes resulting from interactions with the Stoner continuum. 
The neglect of spin-flip excitations implies that Heisenberg mappings effectively correspond to an adiabatic approximation for the magnon dynamics \cite{Halilov1998_adiabatic_spin_dynamics_from_SDFT,Niu1998,Niu1999}, an approximation which is justified whenever the time scale of transverse fluctuations (spin waves) is much longer than for longitudinal fluctuations (spin-flips). This is typically the case for insulators, but is not expected to hold for itinerant magnets in general, except at low spin wave energies. The Heisenberg parameters are crucial for predicting spin wave dispersions, but are also important for other properties. For example they constitute the input to micromagnetic modelling \cite{Bjork2021MagTense} and atomistic spin dynamics \cite{Eriksson2008UppASD,Blugel2019Spirit,Evans2014VampireCode}, 
which can be applied to unravel more complex dynamical effects such as skyrmion motion \cite{Vedmedenko2020Chiral_spin_textures_in_racetrack_devices} and domain wall formation.

The Heisenberg parameters can be computed in real-space from total energy calculations in supercells using various sets of collinear spin configurations \cite{Jin2008T_C_supercell, Olsen2017,Torelli2018Critical_temperatures_in_2D}, or in reciprocal space by the frozen magnon approach, i.e.\ by fitting to spin spiral energies using the generalized Bloch theorem \cite{Halilov1998_adiabatic_spin_dynamics_from_SDFT}. However, the former approach is not applicable to itinerant magnets where long range exchange interactions are expected to play an important role. In addition, the generalized Bloch theorem does not allow for inclusion of spin-orbit coupling, which can only be included perturbatively as a post processing step. Finally, both approaches require calculations of several different spin configurations, which becomes highly impracticable if many parameters are needed for a proper description. As an alternative to the ground state DFT approaches, one can use analytical perturbation theory to express Heisenberg parameters in terms of Green's functions \cite{Rudenko2009Weak_ferromagnetism}, scattering path operators \cite{Liechtenstein87_LSDA_approach_to_ferromagnetic_metals_and_alloys} or linear response functions {\cite{Liechtenstein1995,Wan2006}} that may be calculated directly from the magnetic ground state itself. The idea is to compare the change of energy between the Heisenberg model and DFT calculations under arbitrarily small variations of the spin configuration around the ground state. The process is greatly simplified by the magnetic force theorem (MFT), which states that to linear order, the total energy change in DFT can be calculated from the change in band energy \cite{Liechtenstein87_LSDA_approach_to_ferromagnetic_metals_and_alloys, Mazurenko05_weak_ferromagnetism_in_AFMs}.

Originally the MFT was used for collinear ferromagnets in the context of the Korringa-Kohn-Rostoker (KKR) formalism using the LDA functional \cite{Liechtenstein87_LSDA_approach_to_ferromagnetic_metals_and_alloys}. MFT based formulas were later derived for non-collinear systems \cite{Antropov1997_exch_inter_in_magnets,Antropov1999Aspects_of_spin_dynamics} within KKR, for the Projector Augmented Wave (PAW) method with an LCAO basis \cite{Mazurenko05_weak_ferromagnetism_in_AFMs,Mazurenko14_first_principle_magnetic_excitations,Mazurenko2021A_DMI_guide}, and in terms of Wannier functions expressed in a plane wave basis \cite{Korotin2015Plane-wave-based_LKAG,Nomoto2020Local_force_method_for_tight_binding}. These methods have been extensively tested \cite{Turek06_Heisenberg_model_for_itinerant_magnets,Yoon2018Reliability_of_Magnetic-force_theory}, including comparisons to supercell and frozen magnon calculations \cite{Lezaic2013Exchange_with_local_moment_corrections,Zimmermann2019Comparison_of_ab_init_magnetic_parameters}. The MFT has also been applied to the LDA+U functional \cite{Liechtenstein1995} and in Dynamical Mean-Field Theory (DMFT) \cite{Wan2006} to express Heisenberg parameters in terms of the magnetic susceptibility. More recently, the MFT has been used to map DFT calculations onto tight-binding models \cite{Nomoto2020Local_force_method_for_tight_binding} and to resolve magnetic interactions into individual orbital contributions using localized basis sets \cite{Solovyev2021_exch_inter_and_MFT,Kashin2020Orbitally_resolved_magnetism}.

Here, we present an implementation of the MFT for the computation of Heisenberg parameters using a pure plane wave basis within the PAW formalism \cite{Blochl2003PAW_method}. We start by deriving a basis and lattice independent MFT expression for the full microscopic Heisenberg exchange tensor. The result is expressed in terms of static magnetic susceptibilities and it is shown that our expressions can equivalently be written in terms of Green's functions.
We then calculate spin wave dispersion relations for Fe(bcc), Ni(fcc), Co(fcc) and Co(hcp) and find good agreement with previous calculations.  We emphasize that while the calculations presented here are excluding spin-orbit coupling, the expressions derived for the Heisenberg parameters can in principle be applied to calculations of the full exchange tensor from which e.g. the Dzyaloshinskii-Moriya interations can be extracted.

The paper is organized as follows. In Sec.\ \ref{sec:Theory} we briefly present the formalism of spin-polarized DFT and compare it to a classical microscopic spin model of the magnetic texture based on quadratic spin interactions.
We then apply perturbation theory to show that the spin interactions can be expressed directly in terms of the Kohn-Sham susceptibility tensor. In Sec.\ \ref{sec:Implementation} the plane wave implementation is presented and it is shown that the present formulation allows for complete flexibility in the definition of a Heisenberg lattice model, which is required for extracting spin wave spectra. In Sec.\ \ref{sec:Results_and_discussion} we show the calculated spin wave spectra of bulk Fe, Co and Ni and we discuss limitations of the method by comparing to TDDFT spectra from the literature. Finally, we give a summary of our findings in Sec. \ref{sec:Conclusion}. In addition to the main body of the text, relevant expressions and derivations are supplied in the Appendix. App.\ \ref{appsec:Standard_and_rotated_susceptibility} provides a summary of the four-component susceptibility tensor and App.\ \ref{appsec:Delta_E_DFT} provides a detailed derivation of the microscopic exchange parameters from perturbation theory.
In App. \ref{appsec:Relation_to_Greens_function_results}, the equivalence between our results and the well known Green's function expression for collinear systems is shown and in App.\ \ref{appsec:site_kernels} we provide explicit formulas for the site-kernel in standard geometries.

\section{Theory \label{sec:Theory}}

\subsection{DFT description} \label{sec:DFT model}

When treating spin-polarized systems within DFT, the fundamental quantity of interest is the density matrix
\begin{subequations}
\begin{align}
    &\rho(\mathbf{r}) = \begin{pmatrix}
    n_{\uparrow\uparrow}(\mathbf{r}) & n_{\uparrow\downarrow}(\mathbf{r}) \\ n_{\downarrow\uparrow}(\mathbf{r}) & n_{\downarrow\downarrow}(\mathbf{r})
    \end{pmatrix}
    = \frac{1}{2}\left[n(\mathbf{r}) \sigma_0 + \mathbf{m}(\mathbf{r}) \boldsymbol{\cdot} \boldsymbol{\sigma}\right] \\
    &n(\mathbf{r}) = \mathrm{Tr}\{\rho(\mathbf{r})\}\\
    &\mathbf{m}(\mathbf{r}) = \mathrm{Tr}\{\rho(\mathbf{r})\sigmaVec\} = m(\mathbf{r}) \mathbf{u}(\mathbf{r})\\
    &\boldsymbol{\sigma} = \mathbf{e}_x \sigma_x + \mathbf{e}_y \sigma_y + \mathbf{e}_z \sigma_z 
\end{align}
\label{eq:density_matrix}
\end{subequations}
where $\sigma_\alpha$ are the Pauli matrices, $\sigma_0$ is the identity matrix, $n$ is the electron density and $\mathbf{m}$ is the magnetization with magnitude $m = |\mathbf{m}|$ and direction $\mathbf{u} = \mathbf{m}/m$.
The density matrix $\rho$ can be written in terms of occupied Kohn-Sham orbitals:
\begingroup
\renewcommand*{\arraystretch}{1.5}
\begin{align}
    &\rho(\mathbf{r}) =  \sum_n^{\mathrm{occ.}}\begin{pmatrix}
    |\psi_{n,\uparrow}(\mathbf{r})|^2 & \psi_{n,\uparrow}(\mathbf{r})\psi_{n,\downarrow}^*(\mathbf{r}) 
    \\
    \psi_{n,\downarrow}(\mathbf{r})\psi_{n,\uparrow}^*(\mathbf{r}) & |\psi_{n,\downarrow}(\mathbf{r})|^2
    \end{pmatrix}
    \label{eq:hat_n_from_KS_eigenstates_SO}
\end{align}
\endgroup
where $\psi_{n,s}(\mathbf{r})$ denotes the spin component $s$ of the Kohn-Sham spinor $|\psi_n\rangle$. The Kohn-Sham spinors solve the single-particle Schr\"{o}dinger equation
\begin{align}
    H_\mathrm{KS} |\psi_n\rangle = \varepsilon_n |\psi_n\rangle  \label{eq:Kohn_Sham_problem},
\end{align}
where $\varepsilon_n$ denotes the Kohn-Sham eigenenergies. 

In the Local Density Approximation (LDA), $H_\mathrm{KS}$ may be written as:
\begin{align}\label{eq:H_KS}
    &H_\mathrm{KS}[\rho] = H_0[n, m] + \mathbf{B}^{\mathrm{xc}}[n, \mathbf{m}]\boldsymbol{\cdot} \sigmaVec.
\end{align}
The first term includes scalar-relativistic terms as well as spin-orbit coupling, but depends only on the magnitude of the magnetization $m$ as well as the density. All dependence on the magnetization direction is governed by the effective exchange-correlation magnetic field $\mathbf{B}^{\mathrm{xc}}$, which in the LDA is given by
\begin{align}
 \mathbf{B}^{\mathrm{xc}}(\mathbf{r}) = \frac{\delta E_{\mathrm{xc}}}{\delta m(\mathbf{r})}\mathbf{u}(\mathbf{r})\equiv B^\mathrm{xc}(\mathbf{r})\mathbf{u}(\mathbf{r}),
 \label{eq:LDA_B_xc}
\end{align}
where $B^\mathrm{xc}(\mathbf{r})$ is introduced as the functional derivative of the exchange-correlation energy with respect to the magnitude of the magnetization.

\subsection{Classical spin model description}\label{sec:classical spin model}
In DFT one typically searches for the ground state by minimizing an energy functional $E[n,\mathbf{m}]$. 
Formally, the ground state energy may be written as a functional of only the magnetization direction $E[\mathbf{u}]$ by minimizing the expectation value of the Hamiltonian subject to the constraint that the many-body wavefunction yields $\mathbf{u}$. The ground state density $n$ and magnetization magnitude $m$ are then determined implicitly by $\mathbf{u}$ and the total energy to second order in $\mathbf{u}$ can be written as
\begin{align}
    E_{\mathrm{SM}} = -\frac{1}{2} \iint \mathrm{d}\mathbf{r} \mathrm{d}\mathbf{r}'\, \mathbf{u}^T(\mathbf{r}) \mathrm{J}(\mathbf{r}, \mathbf{r}') \mathbf{u}(\mathbf{r}'),   \label{eq:E_SM_cont}
\end{align}
where matrix multiplication is implied and $\mathrm{J}(\mathbf{r},\mathbf{r}')$ denotes the microscopic exchange tensor with components $J^{\alpha\beta}$ where $\alpha,\beta\in\{x, y, z\}$. 
Since the ground state is stationary with respect to the ground state magnetization direction $\mathbf{u}_0$, Eq. \eqref{eq:E_SM_cont} becomes exact for sufficiently small deviations from the ground state. For time-independent properties, infinitesimal rotations are thus represented by a continuum version of the classical Heisenberg model. However, for dynamical properties (spin wave excitations) the model is relevant only under the assumption that the charge and longitudinal magnetic degrees of freedom follow $\mathbf{u}$ adiabatically.

In matrix form the exchange tensor is written
\begin{align*}
    \mathrm{J}(\mathbf{r}, \mathbf{r}') 
    =
    \begin{pmatrix}
    J^{xx} & J^{xy} & J^{xz} \\
    J^{yx} & J^{yy} & J^{yz} \\
    J^{zx} & J^{zy} & J^{zz}
    \end{pmatrix} \Bigg \vert_{(\mathbf{r}, \mathbf{r}')},
\end{align*}
and one can divide $\mathrm{J}$ into an isotropic, a traceless symmetric and an antisymmetric part,
\begin{align}
    \mathrm{J}(\mathbf{r}, \mathbf{r}') = J(\mathbf{r}, \mathbf{r}')\mathcal{I} + \mathrm{J}^S(\mathbf{r}, \mathbf{r}') + \mathrm{J}^A(\mathbf{r}, \mathbf{r}'),
\end{align}
where
\begin{align}
    &J(\mathbf{r}, \mathbf{r}') = \frac{1}{3}{\mathrm{J}(\mathbf{r}, \mathbf{r}')}, \\
    &\mathrm{J}^S(\mathbf{r}, \mathbf{r}') = \frac{1}{2}[\mathrm{J}(\mathbf{r}, \mathbf{r}') + \mathrm{J}^T(\mathbf{r}, \mathbf{r}')] - J(\mathbf{r}, \mathbf{r}') \mathcal{I}, \\
    &\mathrm{J}^A(\mathbf{r}, \mathbf{r}') = \frac{1}{2}[\mathrm{J}(\mathbf{r}, \mathbf{r}') - \mathrm{J}^T(\mathbf{r}, \mathbf{r}')] \label{eq:J_division}.
\end{align}
The antisymmetric part comprises the Dzyaloshinskii-Moriya vector \cite{Dzyaloshinsky1958a,Moriya1960,Kvashin2020RelaticisticExchangeInteractions}, while the symmetric traceless part gives rise to Kitaev interactions \cite{Xu2018a}. In the absence of spin-orbit coupling only the isotropic part is nonvanishing.

\subsection{Perturbative rotations 
and the magnetic force theorem 
\label{subsec:perturbative_rotation_with_MFT}}

\begin{figure}
    \centering
    \includegraphics[width=\columnwidth]{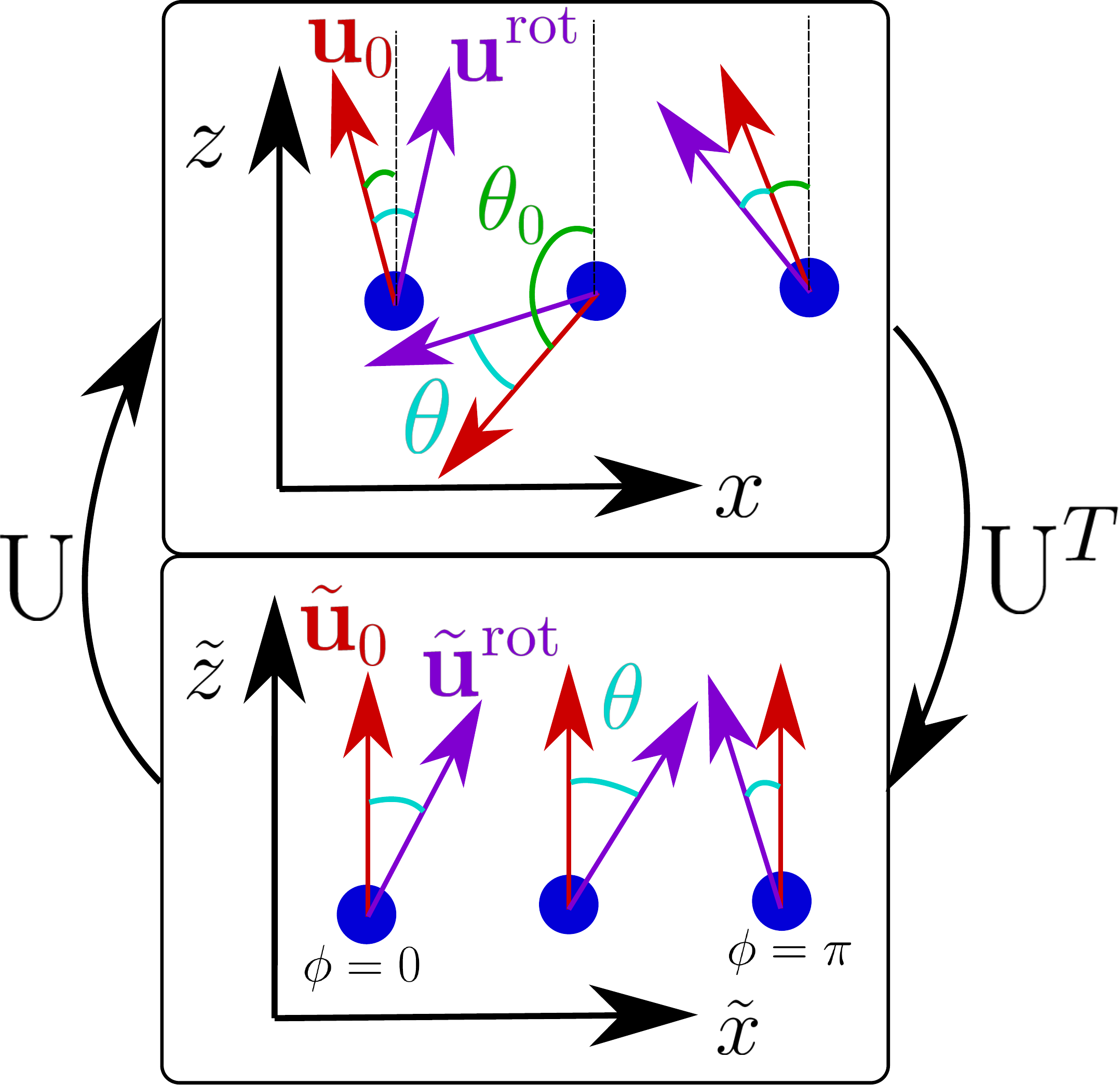}
    \caption{Illustration of the applied coordinate transform. $\phi, \phi_0 \in \{0, \pi\}$ for clarity.
    Top: Laboratory frame with noncollinear ground state, $\mathbf{u}_0$, and perturbed state, $\mathbf{u}^{\mathrm{rot}}$. Bottom: Same state in local coordinates with $\Tilde{\mathbf{u}}_0$ along local $z$-axis.}
    \label{fig:angle_sketch}
\end{figure}

Our main purpose in the following will be to calculate the microscopic exchange tensor in Eq. \eqref{eq:E_SM_cont} from first principles within the LDA. Both the {\it ab initio} electronic structure of Sec. \ref{sec:DFT model} and the classical spin model of Sec. \ref{sec:classical spin model} depend on the vector field $\mathbf{u}(\mathbf{r})$, which gives the local magnetization direction. The ground state magnetic configuration is characterized in terms of $\mathbf{u}_0(\mathbf{r})$, and is allowed to take a noncollinear form in the general case. Inspired by Refs. \cite{Haraldsen2009Spin_rotation_technique,Toth2015LinearSpinWaveTheory}, we consider two reference frames; a global one (the laboratory frame) where $\mathbf{u}_0(\mathbf{r})$ has spherical coordinates $\theta_0(\mathbf{r}), \phi_0(\mathbf{r})$, and a local one (the local frame), with $z$-axis along $\mathbf{u}_0(\mathbf{r})$. The two reference frames are related by a coordinate transform rotating locally the spin direction by $\theta_0(\mathbf{r})$ around $y$ and $\phi_0(\mathbf{r})$ around $z$:
\begin{align}
    \mathrm{U}(\mathbf{r}) = \begin{pmatrix}\cos \phi_0 \cos \theta_0 & -\sin \phi_0 & \cos \phi_0 \sin \theta_0 \\ \sin \phi_0 \cos \theta_0 & \cos \phi_0 & \sin \phi_0 \sin \theta_0 \\ -\sin \theta_0 & 0 & \cos \theta_0 \end{pmatrix}
    \Bigg \vert_{\mathbf{r}}. \label{eq:U(r)}
\end{align}
By applying the inverse mapping $\mathrm{U}^T$ to the magnetic ground state in the usual laboratory frame, the magnetization is mapped from an arbitrary noncollinear configuration onto a ferromagnetic configuration in the local reference frame, greatly simplifying the subsequent algebra. The two reference frames and the coordinate transform between them is illustrated in figure \ref{fig:angle_sketch} for the $xz$-plane. 

In the local frame of reference, the classical spin model \eqref{eq:E_SM_cont} is given by
\begin{align}
    E_{\mathrm{SM}} = -\frac{1}{2} \iint \mathrm{d}\mathbf{r} \mathrm{d}\mathbf{r}'\, \Tilde{\mathbf{u}}^T(\mathbf{r}) \Tilde{\mathrm{J}}(\mathbf{r}, \mathbf{r}') \Tilde{\mathbf{u}}(\mathbf{r}')  \label{eq:E_SM_rot}
\end{align}
where
\begin{equation}
    \Tilde{\mathbf{u}}(\mathbf{r}) = \mathrm{U}^T(\mathbf{r})\mathbf{u}(\mathbf{r}), 
    \label{eq:u_tilde}
\end{equation}
\begin{equation}
    \Tilde{\mathrm{J}}(\mathbf{r}, \mathbf{r}') = \mathrm{U}^T(\mathbf{r})\mathrm{J}(\mathbf{r}, \mathbf{r}')\mathrm{U}(\mathbf{r}'),  \label{eq:J_tilde}
\end{equation}
and $\Tilde{\mathbf{u}}_0(\mathbf{r})$ will, by definition, be a constant unit vector pointing along the $z$-direction. 
The next step is to relate the energy cost of perturbations to the ground state spin direction in the classical spin model to an analogous DFT treatment. Doing so, a DFT expression for the exchange tensor $\Tilde{\mathrm{J}}(\mathbf{r}, \mathbf{r}')$ can be identified. 

Let $\theta(\mathbf{r}), \phi(\mathbf{r})$ be a local set of spherical coordinates corresponding to a given perturbatively rotated configuration $\mathbf{u}^{\mathrm{rot}}(\mathbf{r})$, 
such that $\theta_0, \phi_0$ describe the ground state and $\theta, \phi$ describe rotations \textit{relative} to the ground state, see figure \ref{fig:angle_sketch}.
The polar angle $\theta$ is considered as a perturbative (infinitesimal) quantity ($\phi$ is allowed to vary freely) and orders in $\theta$ are separated as follows,
\begin{align}
    \mathbf{u}^{\mathrm{rot}} = \mathbf{u}_0 + \delta \mathbf{u} + \delta^2 \mathbf{u} + \mathcal{O}(\theta^3),
\end{align}
where
\begin{align}
    \Tilde{\mathbf{u}}_0 = \begin{pmatrix}0\\0\\1\end{pmatrix}, \quad \delta \Tilde{\mathbf{u}} = \theta\begin{pmatrix}\cos\phi \\ \sin\phi \\ 0\end{pmatrix}, \quad \delta^2 \Tilde{\mathbf{u}} = -\frac{1}{2}\theta^2 \Tilde{\mathbf{u}}_0.
    \label{eq:rotated directions, ordered}
\end{align}
Changes to the energy enter at second order in $\theta$, since first order variations in $\theta$ correspond to a finite torque \cite{Mazurenko05_weak_ferromagnetism_in_AFMs}, which is absent when perturbing around the ground state.  

In the classical spin model \eqref{eq:E_SM_rot}, the change in energy to second order in $\theta$ is given by
\begin{align}
    \delta^2 E_{\mathrm{SM}} &= \frac{1}{2} \int \mathrm{d}\mathbf{r}\, \theta(\mathbf{r})^2 \int \mathrm{d}\mathbf{r}'\, \Tilde{J}^{zz}(\mathbf{r}, \mathbf{r}')
    \notag\\
    &\quad -\frac{1}{2}\iint \mathrm{d}\mathbf{r} \mathrm{d}\mathbf{r}'\, \delta \Tilde{\mathbf{u}}^T(\mathbf{r}) \Tilde{\mathrm{J}}(\mathbf{r}, \mathbf{r}') \delta \Tilde{\mathbf{u}}(\mathbf{r}') \label{eq:Delta_E^2_SM_continuous}
\end{align}
where it was used that $J^{zz}(\mathbf{r},\mathbf{r}') = J^{zz}(\mathbf{r}', \mathbf{r})$. In the DFT treatment, one may perform calculations with the spin direction constrained to comply with $\mathbf{u}^{\mathrm{rot}}(\mathbf{r})$. The total energy can then be written as $E_{\mathrm{DFT}} = E_{\mathrm{SP}} + E_{\mathrm{rest}}$, where $E_{\mathrm{SP}} = \sum_n^{\mathrm{occ}} \varepsilon_n$ is the sum of occupied single-particle energies and $E_{\mathrm{rest}}$ encodes all the additional terms. In the LDA, $E_{\mathrm{rest}}$ does not depend explicitly on $\mathbf{u}$, but a local rotation could still induce a relaxation of charge and spin densities, i.e.\ $n_0,m_0 \xrightarrow{} n_0 + \delta n, m_0 + \delta m$, where $n_0,m_0$ are the ground state values. Nevertheless, the magnetic force theorem (MFT) dictates that  $\delta n$ and $\delta m$ do not change $E_{\mathrm{DFT}}$ to first order, i.e.\
\begin{align}
    \Delta E_{\mathrm{DFT}} &= E_{\mathrm{SP}}[\mathbf{u}^{\mathrm{rot}}, n_0, m_0] - E_{\mathrm{SP}}[\mathbf{u}_0, n_0, m_0]\notag \\
    &\quad+ \mathcal{O}[\delta n^2, \delta m^2]    \label{eq:MFT}.
\end{align}
The MFT was originally shown explicitly under the assumption of the LDA \cite{Liechtenstein87_LSDA_approach_to_ferromagnetic_metals_and_alloys}, but was later generalized to any functional satisfying gauge symmetry constraints \cite{Solovyev1998EffectiveSingleParticlePotentialsMnO}. 
In addition, it can be shown that only the first order changes in density and magnetization are relevant to leading (second) order in the polar rotation angle $\theta$ \cite{Liechtenstein87_LSDA_approach_to_ferromagnetic_metals_and_alloys,Solovyev1998EffectiveSingleParticlePotentialsMnO}. Thus, when comparing the DFT treatment to the leading order perturbation in the classical spin model \eqref{eq:Delta_E^2_SM_continuous}, only changes in the single-particle energies need to be considered and 
\begin{align}
    \Delta E_{\mathrm{DFT}} &\simeq E_{\mathrm{SP}}[\mathbf{u}^{\mathrm{rot}}, n_0, m_0] - E_{\mathrm{SP}}[\mathbf{u}_0, n_0, m_0]
    \label{eq:single-particle E_dif}
\end{align}
including all second order terms in $\theta$.

\subsection{Changes in the single-particle energies \label{subsec:change_in_single-particle_energies}}

To complete the comparison of perturbative rotations in the DFT and classical spin model descriptions, the final step is to identify the change in single-particle energies \eqref{eq:single-particle E_dif} to second order in $\theta$.
Since  $H_0=H_0[n,m]$ in Eq. \eqref{eq:H_KS} has no explicit dependence on $\mathbf{u}$, it is invariant under rotations of the local spin direction (neglecting relaxation effects according to Eq.\ \eqref{eq:single-particle E_dif}). Thus the only part of $H_{\mathrm{KS}}$ which changes is $\mathbf{B}^{\mathrm{xc}}$.
In the LDA $\mathbf{B}^{\mathrm{xc}}$ is parallel with $\mathbf{u}$: $\mathbf{B}^{\mathrm{xc}}\boldsymbol{\cdot}\sigmaVec=B^{\mathrm{xc}}\mathbf{u}\boldsymbol{\cdot}\sigmaVec$ and
the relevant change in $H_{\mathrm{KS}}$ under a general rotation can thus be written
\begin{align}
    \Delta H_{\mathrm{KS}}
    &\simeq B^{\mathrm{xc}}(\mathbf{u}^{\mathrm{rot}} - \mathbf{u}^0) \boldsymbol{\cdot} \sigmaVec\notag\\
    &= B^{\mathrm{xc}} \big[\mathrm{U} (\Tilde{\mathbf{u}}^{\mathrm{rot}} - \Tilde{\mathbf{u}}^0)\big]^T \sigmaVec\notag\\
    &= B^{\mathrm{xc}} (\Tilde{\mathbf{u}}^{\mathrm{rot}} - \Tilde{\mathbf{u}}^0)^T \mathrm{U}^T \sigmaVec\notag\\
    &= B^{\mathrm{xc}} (\Tilde{\mathbf{u}}^{\mathrm{rot}} - \Tilde{\mathbf{u}}^0) \boldsymbol{\cdot} \Tilde{\sigmaVec},
    \label{eq:H_KS under rotation 1}
\end{align}
where matrix multiplication is implied in the middle two equations and $\Tilde{\sigmaVec}=\Tilde{\sigmaVec}(\mathbf{r})$ is the locally rotated Pauli vector
\begin{align}
    \Tilde{\sigmaVec}(\mathbf{r}) = \mathrm{U}^T(\mathbf{r}) \sigmaVec = \mathrm{U}_{1/2}^{\dag}(\mathbf{r}) \sigmaVec \mathrm{U}_{1/2}(\mathbf{r}).
    \label{eq:locally rotated pauli vector}
\end{align}
Here, $\mathrm{U}_{1/2}$ is the spinor rotation operator corresponding to the vector rotation $\mathrm{U}$:
\begin{align}
    \mathrm{U}_{1/2}(\mathbf{r}) = \begin{pmatrix}\ee^{i\phi_0/2} \cos(\theta_0/2) & \ee^{-i\phi_0/2} \sin(\theta_0/2) \\ -\ee^{i\phi_0/2} \sin(\theta_0/2) & \ee^{-i\phi_0/2} \cos(\theta_0/2)\end{pmatrix}\bigg \vert_{\mathbf{r}}. \label{eq:hat_U(r)}
\end{align}
Finally, expanding Eq. \eqref{eq:H_KS under rotation 1} to second order in $\theta$,
\begin{align}
    \Delta H_{\mathrm{KS}} 
    &=\underbrace{B^{\mathrm{xc}} \delta \Tilde{\mathbf{u}} \boldsymbol{\cdot} \Tilde{\sigmaVec}}_{\delta H_{\mathrm{KS}}} \underbrace{-\frac{1}{2}\theta^2 B^{\mathrm{xc}} \mathbf{e}_z \boldsymbol{\cdot} \Tilde{\sigmaVec}}_{\delta^2 H_{\mathrm{KS}}} + \mathcal{O}(\theta^3).
\end{align}
The change in Kohn-Sham eigenenergies can then be evaluated using standard second order perturbation theory:
\begin{align*}
    \Delta\varepsilon_n = \langle\psi_n| \Delta H_{\mathrm{KS}} |\psi_n\rangle + \sum_{m\neq n} \frac{|\langle\psi_m|\Delta H_{\mathrm{KS}} |\psi_n\rangle^2|}{\varepsilon_n - \varepsilon_m}.
\end{align*}
Thus, we obtain a second order change in the total energy given by
\begin{align}
   \Delta E_{\mathrm{DFT}}^{(2)} = \sum_n^{\mathrm{occ}}\Bigg[ &\langle\psi_n| \delta^2 H_{\mathrm{KS}} |\psi_n\rangle 
   \notag\\
   &+ \sum_{m\neq n} \frac{\big|\langle\psi_m|\delta H_{\mathrm{KS}} |\psi_n\rangle\big|^2}{\varepsilon_n - \varepsilon_m}\Bigg], \label{eq:Delta_E_DFT}
\end{align}
which may be compared to the analogous result for the classical spin model \eqref{eq:Delta_E^2_SM_continuous} in order to identify $\Tilde{\mathrm{J}}(\mathbf{r}, \mathbf{r}')$.

\subsection{MFT relations between the microscopic exchange tensor and the Kohn-Sham susceptibility}
After having introduced the four-component susceptibility tensor in App. \ref{appsec:Standard_and_rotated_susceptibility}, the expression \eqref{eq:Delta_E_DFT} is evaluated in terms of the susceptibility in App.  \ref{appsec:Delta_E_DFT}. Through comparison with Eq. \eqref{eq:Delta_E^2_SM_continuous} we obtain
\begin{align}
    \int \mathrm{d}\mathbf{r}'\, \Tilde{J}^{zz}(\mathbf{r}, \mathbf{r}')  = -B^{\mathrm{xc}}(\mathbf{r}) m_0(\mathbf{r}), \label{eq:Jzz_from_magnetisation}
\end{align}
and
\begin{align}
    & \Tilde{J}^{\alpha\beta}(\mathbf{r}, \mathbf{r}') = -B^{\mathrm{xc}}(\mathbf{r}) \mathrm{Re}\left\{\Tilde{\chi}'^{\alpha\beta}_{\mathrm{KS}}(\mathbf{r}, \mathbf{r}')\right\} B^{\mathrm{xc}}(\mathbf{r}'),    \label{eq:J_cartesian_tilde}
\end{align}
where $\alpha,\beta\in\{x,y\}$ in Eq. \eqref{eq:J_cartesian_tilde}. The tensor $\Tilde{\chi}_{\mathrm{KS}}'^{\alpha\beta}(\mathbf{r},\mathbf{r}')$
denotes the reactive part of the static cartesian Kohn-Sham  susceptibility tensor in the local frame of reference, which is related by a simple basis change to the susceptibility tensor in the laboratory frame, see Eq. \eqref{eq:chi_rotated}. It is straightforward to verify that $\Tilde{\chi}_{\mathrm{KS}}'^{\alpha\beta}$ is hermitian in the sense that $\Tilde{\chi}'^{\beta\alpha*}_{\mathrm{KS}}(\mathbf{r}', \mathbf{r})=\Tilde{\chi}'^{\alpha\beta}_{\mathrm{KS}}(\mathbf{r}, \mathbf{r}')$ (see \cref{appsec:Delta_E_DFT}).
Furthermore, if two points $\mathbf{r}_1$ and $\mathbf{r}_2$ are related by inversion symmetry then $\Tilde{\chi}'^{xy}_{\mathrm{KS}}(\mathbf{r}_1, \mathbf{r}_2)=\Tilde{\chi}'^{xy}_{\mathrm{KS}}(\mathbf{r}_2, \mathbf{r}_1)$, implying that $\Tilde{J}^{xy}(\mathbf{r}_1, \mathbf{r}_2)=\Tilde{J}^{yx}(\mathbf{r}_1, \mathbf{r}_2)$. This leads to the well-known fact that the Dzyaloshinskii-Moriya interaction (anti-symmetric part of the exchange tensor) vanishes between points related by inversion symmetry \cite{Moriya73_spin_fluctuations}.

Application of Eq. \eqref{eq:J_cartesian_tilde} only yields 4 out of 9 components of $\Tilde{\mathrm{J}}$, but one may in principle repeat the calculation placing the magnetization along the local $x$ and $y$ axes rather than $z$, in order to obtain the full microscopic exchange tensor \cite{Kvashin2020RelaticisticExchangeInteractions,Udvardi2003Ab_init_spin_waves_in_thin_films}. Thus, Eq. \eqref{eq:J_cartesian_tilde} may be formally generalized to include $z$-components, if one bears in mind that the corresponding susceptibilities should be calculated with respect to different magnetization directions. This procedure requires that the susceptibilities calculated with rotated magnetization should be based on non-self-consistent calculations, since otherwise the rotation matrix ($\mathrm{U}$) may change and $\tilde J^{xz}$, for example, will not transform in the same way as $\tilde J^{xy}$. It is not clear at present if this point raises fundamental issues with the approach since the MFT requires a self-consistent point of reference. Nevertheless, the change in the density matrix under a (large) rotation of the ground state magnetization is typically negligible compared to the change in eigenvalues, which in turn dominates the change in the susceptibility.
This suggests that one can indeed calculate the full microscopic exchange tensor according to the procedure described above followed by a transformation back to the laboratory frame:
\begin{align}
    \mathrm{J}(\mathbf{r}, \mathbf{r}') = \mathrm{U}(\mathbf{r})\Tilde{\mathrm{J}}(\mathbf{r}, \mathbf{r}')\mathrm{U}^T(\mathbf{r}').  \label{eq:J_tilde_to_lab}
\end{align}
The calculation is largely simplified in collinear systems where $\theta_0\in \{0,\pi\}$, in which case the rotation matrix $\mathrm{U}$ becomes block diagonal and the $x$,$y$-components do not mix with the $z$-component\cite{DosSantosDias2015}, see Eq. \eqref{eq:U(r)}. Thus, the transformation \eqref{eq:J_tilde_to_lab} can be applied directly to the $x$,$y$-components in Eq. \eqref{eq:J_cartesian_tilde} to yield the transverse components of the microscopic exchange tensor in terms of the susceptibility in the laboratory frame,
\begin{align}
    &J^{\alpha\beta}(\mathbf{r}, \mathbf{r}') = -B^{\mathrm{xc}}(\mathbf{r}) \mathrm{Re}\left\{\chi'^{\alpha\beta}_{\mathrm{KS}}(\mathbf{r}, \mathbf{r}')\right\} B^{\mathrm{xc}}(\mathbf{r}')   \label{eq:J^cartesian_tilde},
\end{align}
where $\alpha,\beta\in\{x,y\}$.

\subsection{Collinear magnets without spin-orbit coupling}\label{sec:theory_simple_ferromagnet}
In the absence of spin-orbit coupling, the spin-polarization along the $z$-axis can be taken as a good quantum number and the Kohn-Sham eigenstates can be written with only a single nonvanishing spin component.
As discussed in App. \ref{appsec:Standard_and_rotated_susceptibility}, the susceptibility tensor is then significantly simplified with $\chi'^{xy}_{\mathrm{KS}}(\mathbf{r}, \mathbf{r}')=\chi'^{yx}_{\mathrm{KS}}(\mathbf{r}, \mathbf{r}') = 0$ and $\chi'^{xx}_\mathrm{KS}=\chi'^{yy}_\mathrm{KS}=2\chi'^{+-}_\mathrm{KS}$ where
\begin{align}
    \chi'^{+-}_{\mathrm{KS}}(\mathbf{r},\mathbf{r}') =  \sum_{n,m} &\frac{f_{n\uparrow} - f_{m\downarrow}}{\varepsilon_{n\uparrow} - \varepsilon_{m\downarrow}} \psi_{n\uparrow}^*(\mathbf{r}) \psi_{m\downarrow}(\mathbf{r}) 
    \psi_{m\downarrow}^*(\mathbf{r}')
    \psi_{n\uparrow}(\mathbf{r}'). \label{eq:chi^+-_KS}
\end{align}
Here, the spin has been included explicitly in the Kohn-Sham state index $\psi_n\rightarrow\psi_{ns}$ where $s\in\{\uparrow, \downarrow\}$ and $n$ now denotes the remaining quantum numbers.
In addition, without spin-orbit coupling the spin direction of the ground state is arbitrary, meaning that the exchange tensor becomes completely isotropic: $\mathrm{J}=J\mathcal{I}$. Thus, based on Eq. \eqref{eq:J^cartesian_tilde} it becomes exceedingly simple to compute the microscopic exchange from first principles
\begin{align}
    J(\mathbf{r}, \mathbf{r}') = -2 B^{\mathrm{xc}}(\mathbf{r}) \chi'^{+-}_{\mathrm{KS}}(\mathbf{r}, \mathbf{r}') B^{\mathrm{xc}}(\mathbf{r}') \label{eq:exchange_in_ferromagnet},
\end{align}
with $\chi'^{+-}_\mathrm{KS}$ being a real function since the Kohn-Sham orbitals may be chosen as real when spin-orbit coupling is not included. Lastly, the consistency with Eq. \eqref{eq:Jzz_from_magnetisation} can be verified by noting that $\chi'^{+-}_\mathrm{KS}$ is subject to the sum rule \cite{Katsnelson_2004_Magnetic_susceptibility_LSDA}
\begin{align}
    m_0(\mathbf{r}) = 2 \int \mathrm{d}\mathbf{r}'\, \chi'^{+-}_{\mathrm{KS}}(\mathbf{r}, \mathbf{r}') B^{\mathrm{xc}}(\mathbf{r}') \label{eq:m_0_identity},
\end{align}
implying that Eq. \eqref{eq:Jzz_from_magnetisation} is satisfied with $J^{zz}=J$ as given by Eq. \eqref{eq:exchange_in_ferromagnet}. In \cref{appsec:Relation_to_Greens_function_results}, Eq. \eqref{eq:exchange_in_ferromagnet} is recast in terms of Green's functions, showcasing the equivalence to the expression of Ref. \cite{Bruno03_renormalised_magnetic_force_theorem}. We also note that Eq. \eqref{eq:exchange_in_ferromagnet} comprises the {\it ab initio} generalization of a similar result that has previously been derived in the framework of Hubbard models \cite{Liu1977,Prange1979,Wang1982}.

\section{Implementation \label{sec:Implementation}}

\subsection{Discretization of the classical spin model \label{sec:Discretization}}

\begin{figure}[bt]
    \centering
    \includegraphics[width=0.3\textwidth]{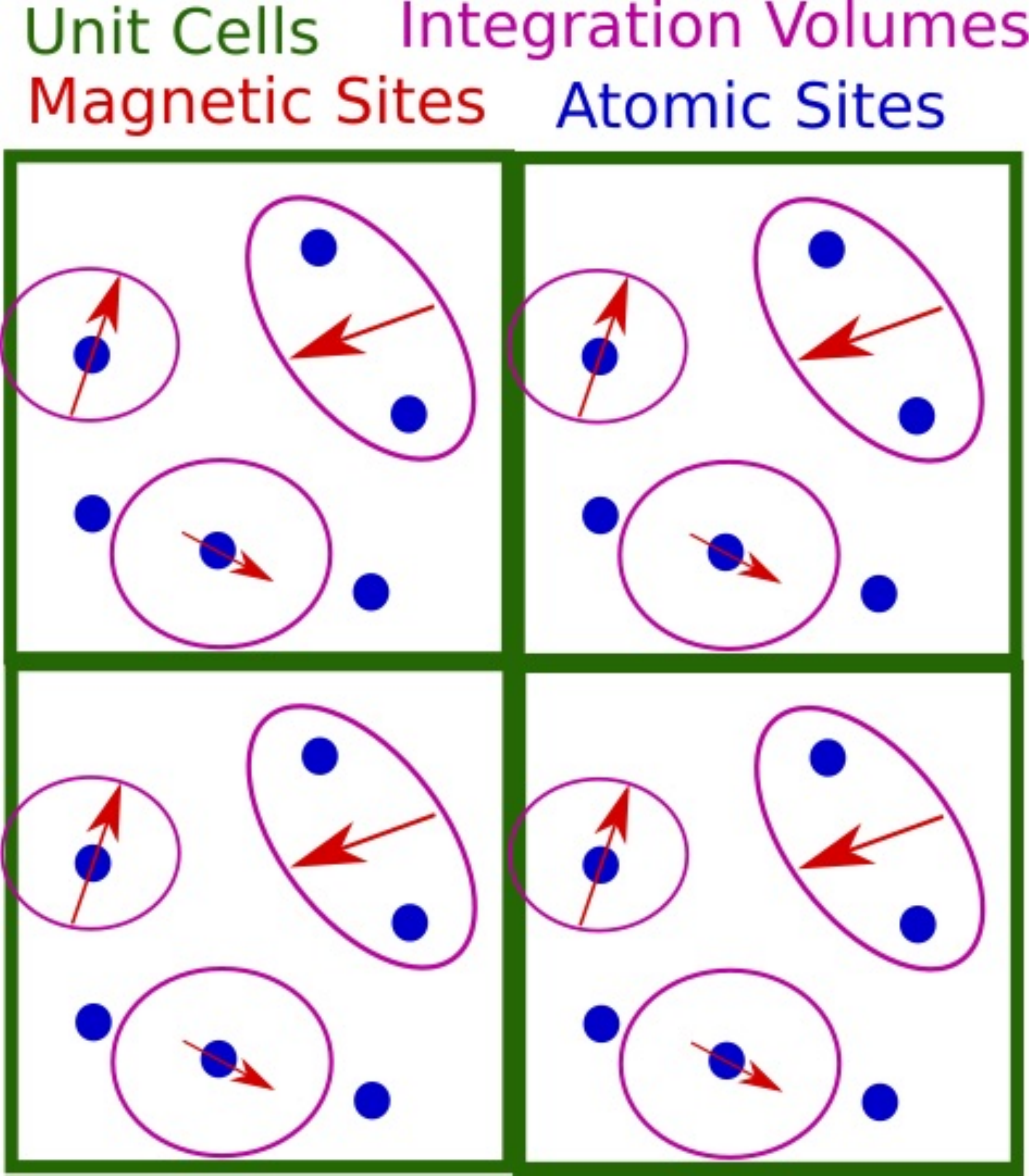}
    \caption{Conceptual sketch of a magnetic material with integration volumes enclosing the magnetic atoms.}
    \label{fig:integration_spheres_on_lattice}
\end{figure}

In the preceding section, the exchange tensor $\mathrm{J}(\mathbf{r}, \mathbf{r}')$, which parametrizes the continuous spin model \eqref{eq:E_SM_cont}, was related to the transverse magnetic susceptibility, see Eqs. \eqref{eq:J_cartesian_tilde}, \eqref{eq:J^cartesian_tilde} and \eqref{eq:exchange_in_ferromagnet}, resulting in a general and basis-independent formulation of the exchange interaction. However, in order to simplify the subsequent analysis using linear spin wave theory, it is convenient to map the problem onto a classical localized spin model of the form 
\begin{align}
E_{\mathrm{LSM}} = -\frac{1}{2} \sum_{\mathbf{R},\mathbf{R}'} \mathbf{u}^T_{\mathbf{R}} \mathrm{J}_{\mathbf{RR}'} \mathbf{u}_{\mathbf{R}'},
    \label{eq:E_LSM}
\end{align}
where $\mathbf{u}_{\mathbf{R}}$ is a unit vector representing the direction of the magnetic moment at site $\mathbf{R}$. Formally, the magnetic sites of the model are defined by choosing a set of appropriate non-overlapping integration volumes $\Omega_{\mathbf{R}}$ enclosing the magnetic atoms of the material, see figure \ref{fig:integration_spheres_on_lattice}. The localized spin model \eqref{eq:E_LSM} is then obtained from the continuous model \eqref{eq:E_SM_cont} by employing the rigid spin approximation,
\begin{equation}
    \mathbf{u}(\mathbf{r}) \simeq \sum_{\mathbf{R}} \Theta(\mathbf{r}\in\Omega_{\mathbf{R}}) \mathbf{u}_{\mathbf{R}},
    \label{eq:rigid_spin_approx}
\end{equation}
where $\Theta(\mathbf{r}\in\Omega_{\mathbf{R}})$ denotes the unit step function, nonzero only for $\mathbf{r}\in\Omega_{\mathbf{R}}$, and $\mathbf{u}_{\mathbf{R}}$ is formally defined as the average direction of the magnetization within the integration volume $\Omega_{\mathbf{R}}$.

For collinear magnets in the absence of spin-orbit coupling (see Sec. \ref{sec:theory_simple_ferromagnet}), only the isotropic part of the exchange tensor is nonzero, i.e.\ $\mathrm{J}_{\mathbf{RR}'}=J_{\mathbf{RR}'}\mathcal{I}$. Inserting the rigid spin approximation \eqref{eq:rigid_spin_approx} into the continuous spin model \eqref{eq:E_SM_cont} and using Eq. \eqref{eq:exchange_in_ferromagnet}, one obtains
\begin{align}
    J_{\mathbf{RR}'}
    &= -2\int_{\Omega_{\mathbf{R}}} \mathrm{d}\mathbf{r} \int_{\Omega_{\mathbf{R}'}} \mathrm{d}\mathbf{r}'\, B^{\mathrm{xc}}(\mathbf{r}) \chi_{\mathrm{KS}}'^{+-}(\mathbf{r}, \mathbf{r}') B^{\mathrm{xc}}(\mathbf{r}')
    \label{eq:J_RR'}.
\end{align}
The rigid spin approximation \eqref{eq:rigid_spin_approx} entails neglecting energy contributions from interstitial regions between the magnetic sites, as well as noncollinear structure within any individual integration volume $\Omega_{\mathbf{R}}$. The latter may seem like an obvious approximation for a ferromagnet, but when analyzing the spin dynamics of the system, also noncollinear configurations in near proximity to the ground state need to be well described. In this sense, the rigid spin approximation implicitly assumes that the direction of the magnetization inside any given magnetic site can be taken as constant on the time scale of the spin dynamics. 

While it is often natural to choose $\mathbf{R}$ as the positions of individual atoms and $\Omega_{\mathbf{R}}$ as spherical regions around the atoms, it should be emphasized that this is not necessary, nor always optimal. 
For instance, Le$\mathrm{\Breve{z}}$ai$\mathrm{\Grave{c}}$ et al.\ considered systems where the exchange interaction between strongly magnetic atoms is renormalized by coupling to surrounding weakly magnetic atoms \cite{Lezaic2013Exchange_with_local_moment_corrections}. 
In that case, or for closely spaced magnetic atoms with exchange interactions that are orders of magnitudes larger than the remaining interactions, assigning multiple atoms to one magnetic site can be reasonable. 
We note that $J_{\mathbf{RR}'}$ may alternatively be evaluated from an energy mapping approach where different spin configurations are mapped to the Heisenberg model. In that case, the predicted values of $J_{\mathbf{RR}'}$ depend on whether the configurations are mapped to a classical or a quantum mechanical model \cite{Torelli2020}. When considering infinitesimal rotations, however, the classical quadratic model is formally exact in the adiabatic limit. This implies that higher order spin interactions may be required for an accurate classical description of general rotations, even if the quantum model is accurate with quadratic interactions only.

\subsection{Isotropic exchange parameters in a plane wave basis \label{subsec:isotropic_exchange_in_reciprocal_space}}

When evaluating the Heisenberg exchange parameters \eqref{eq:J_RR'}, one can freely choose a suitable basis representation for the susceptibility and exchange-correlation magnetic field. Here, we present a plane wave implementation based on the transverse magnetic susceptibility module \cite{Skovhus2021Published_version} in the GPAW electronic structure code \cite{Mortensen2005GPAW,Enkovaara2010GPAW_review}. As the plane wave basis facilitates an, in principle, complete representation of the continuous exchange tensor, it is especially relevant for benchmarking purposes.

Assuming a periodic crystal with reciprocal lattice vectors $\mathbf{G}$, the plane wave representations of $B^{\mathrm{xc}}$ and $\chi'^{+-}_{\mathrm{KS}}$ are given by (see Ref. \cite{Skovhus2021Published_version} for definitions of the Fourier transforms)
\begin{align}
    B^{\mathrm{xc}}(\mathbf{r}) = \frac{1}{\Omega_{\mathrm{cell}}} \sum_{\mathbf{G}} B^{\mathrm{xc}}_{\mathbf{G}} \ee^{i\mathbf{G} \boldsymbol{\cdot} \mathbf{r}},  \label{eq:B^xc_fourier}
\end{align}
and
\begin{align}
    \chi'^{+-}_{\mathrm{KS}}(\mathbf{r}, \mathbf{r}') = \frac{1}{\Omega} \sum_{\mathbf{G,G'}}\sum_{\mathbf{q}}\ee^{i(\mathbf{G} + \mathbf{q})\boldsymbol{\cdot} \mathbf{r}} \chi'^{+-}_{\mathrm{KS}, \mathbf{GG'}}(\mathbf{q}) \ee^{-i(\mathbf{G}' + \mathbf{q}) \boldsymbol{\cdot} \mathbf{r}'}, \label{eq:chi_fourier}
\end{align}
where $\Omega_{\mathrm{cell}}$ is the unit cell volume, $\Omega = N_k\Omega_{\mathrm{cell}}$ is the crystal volume, $N_k$ the number of unit cells and $\mathbf{q}$ denotes the wave vectors within the first Brillouin zone.
It should be noted that $B^{\mathrm{xc}}(\mathbf{r})=B^{\mathrm{xc}}(\mathbf{r}+\mathbf{R})$ and $\chi'^{+-}_{\mathrm{KS}}(\mathbf{r}, \mathbf{r}')=\chi'^{+-}_{\mathrm{KS}}(\mathbf{r}+\mathbf{R}, \mathbf{r}'+\mathbf{R})$ are periodic on the Bravais lattice of the crystal, implying also that the linear response encoded by the susceptibility is diagonal in the reduced wave vector $\mathbf{q}$.

For a Heisenberg model with multiple magnetic sites in the unit cell, the system is naturally divided into magnetic sublattices. To fully utilize the translational symmetry, $i,j$ are introduced as cell indices and $a,b$ as sublattice indices, writing
\begin{align*}
    \mathbf{R} \xrightarrow{} \mathbf{R}_{ia} = \mathbf{R}_i + \boldsymbol{\tau}_{a},
\end{align*}
where $\mathbf{R}_i$ denotes the origin of the $i$'th unit cell and $\boldsymbol{\tau}_{a}$ is the position of the $a$'th magnetic site within a unit cell.
Inserting Eqs. \eqref{eq:B^xc_fourier} and \eqref{eq:chi_fourier} into Eq. \eqref{eq:J_RR'}, and using that $B^{\mathrm{xc}}(\mathbf{r})$ is a real function,
\begin{align}
    J^{ab}_{ij} &= J_{\mathbf{R}_{ia}\mathbf{R}_{jb}}
    \notag\\
    &= - \frac{2}{\Omega} \sum_{\mathbf{G}_1,\mathbf{G}_2,\mathbf{G}_3,\mathbf{G}_4} \sum_{\mathbf{q}} \ee^{i \mathbf{q} \boldsymbol{\cdot} (\mathbf{R}_i - \mathbf{R}_j) } B^{\mathrm{xc}*}_{\mathbf{G}_1} K^{a}_{\mathbf{G}_1\mathbf{G}_2}(-\mathbf{q}) \notag\\
    &\hspace{89pt} \times \chi'^{+-}_{\mathrm{KS},\mathbf{G}_2\mathbf{G}_3}(\mathbf{q}) K^{b}_{\mathbf{G}_3\mathbf{G}_4}(\mathbf{q})B^{\mathrm{xc}}_{\mathbf{G}_4},    \label{eq:J^munu_nn'}
\end{align}
where we have introduced the sublattice site-kernel
\begin{equation}
    K_{\mathbf{G}\mathbf{G}'}^a(\mathbf{q})= \frac{1}{\Omega_{\mathrm{cell}}} \int_{\Omega_{a}} \mathrm{d}\mathbf{r}\, \ee^{-i(\mathbf{G} - \mathbf{G}' + \mathbf{q}) \boldsymbol{\cdot} \mathbf{r}},
    \label{eq:site_kernels}
\end{equation}
and $\Omega_a=\Omega_{\mathbf{R}_{0a}}$ is the integration volume of sublattice $a$ (centered at $\boldsymbol{\tau}_{a}$). 
In App. \ref{appsec:site_kernel_formulas}, we provide analytic formulas for the integral above for spherical, cylindrical and parallelepipedic integration volumes centered at the origin.

\subsection{Lattice Fourier transform of the Heisenberg exchange parameters}

In periodic crystals, the exchange tensor is invariant under lattice translations, $J(\mathbf{r},\mathbf{r}')=J(\mathbf{r}+\mathbf{R}, \mathbf{r}'+\mathbf{R})$, and has a diagonal representation in the reduced wave vector $\mathbf{q}$ as a result. This implies that the Heisenberg exchange parameters \eqref{eq:J^munu_nn'} can be written
\begin{equation}
    J_{ij}^{ab} = \frac{1}{N_k}\sum_{\mathbf{q}} e^{i\mathbf{q}\boldsymbol{\cdot}(\mathbf{R}_{i} - \mathbf{R}_{j})} \bar{J}^{ab}(\mathbf{q}),
    \label{eq:Heisenberg params in terms of Jbar}
\end{equation}
where
\begin{equation}
    \bar{J}^{ab}(\mathbf{q}) = \sum_i J_{0i}^{ab} e^{i\mathbf{q} \boldsymbol{\cdot} \mathbf{R}_i}.
\end{equation}
The energy of frozen spin wave configurations in the classical Heisenberg model \cite{Halilov1998_adiabatic_spin_dynamics_from_SDFT} as well as the spin wave dispersion (see the following section) are directly determined by $\bar{J}^{ab}(\mathbf{q})$. 
Comparing Eqs. \eqref{eq:J^munu_nn'} and \eqref{eq:Heisenberg params in terms of Jbar}, $\bar{J}^{ab}(\mathbf{q})$ may be written in the plane wave basis,
\begin{equation}
    \bar{J}^{ab}(\mathbf{q}) = - \frac{2}{\Omega_{\mathrm{cell}}}  B^{\mathrm{xc}\dagger}K^{a\dagger}(\mathbf{q})\chi'^{+-}_{\mathrm{KS}}(\mathbf{q}) K^{b}(\mathbf{q})B^{\mathrm{xc}},
    \label{eq:J_breve_formula}
\end{equation}
where matrix/vector multiplication in $\mathbf{G}$-indices is implied.

Usually, the onsite exchange elements $J^{aa}_{00}$ are excluded from isotropic Heisenberg models and the lattice Fourier transform is introduced relative to the onsite exchange \cite{Yosida1996}:
\begin{equation}
    J^{ab}(\mathbf{q}) = \bar{J}^{ab}(\mathbf{q}) - J_{00}^{aa}\delta^{ab}.
    \label{eq:J^munu_fourier_transform}
\end{equation}
These definitions are somewhat arbitrary as the onsite exchange elements do not influence the spin wave dispersion. However, in the presence of spin-orbit coupling, the difference in onsite exchange elements along different directions plays a crucial role for the thermodynamic stability of the magnetic ground state, typically formulated in terms of single-ion anisotropy terms in the localized spin model. 
The isotropic onsite exchange, $J^{aa}_{00}$, is itself given by the average of $\bar{J}^{aa}(\mathbf{q})$ over the first Brillouin zone, see Eq. \eqref{eq:Heisenberg params in terms of Jbar}:
\begin{align}
    J^{aa}_{00} = \frac{1}{N_k}\sum_{\mathbf{q}} \bar{J}^{aa}(\mathbf{q}). \label{eq:constant_offset}
\end{align}
Thus, it is necessary to sample the entire Brillouin zone in order to compute the exchange parameters $J^{ab}(\mathbf{q})$. Because the isotropic onsite exchange cancels out in the spin wave dispersion, it is hence generally advantageous to compute the excitation spectra directly from $\bar{J}^{ab}(\mathbf{q})$.

\subsection{Spin wave spectra \label{sec:spin-wave spectra}}
For the isotropic Heisenberg model with ferromagnetic exchange, the magnon energies $\hbar \omega_\alpha(\mathbf{q})$ are given in linear spin wave theory as the eigenvalues of the matrix \cite{Toth2015LinearSpinWaveTheory}
\begin{equation}
    H^{ab}(\mathbf{q}) = \frac{g\mu_{\mathrm{B}}}{\sqrt{M_aM_b}} \left[\sum_c \bar{J}^{ac}(\mathbf{0})\delta_{ab} - \bar{J}^{ab}(\mathbf{q})\right],
    \label{eq:H^munu_q}
\end{equation}
where $\mu_{\mathrm{B}}$ is the Bohr magneton, $g \approx 2$ is the electron g-factor, $M_a$ is the magnitude of the magnetic moment of sublattice $a$ and $\bar{J}$ can be interchanged with $J$ as long as it is done in all terms. The spin wave eigenvalue problem can also be derived in alternative ways, see e.g. Refs. \cite{Halilov1998_adiabatic_spin_dynamics_from_SDFT,kubler2009theory}. In particular, these authors highlight that when modelling the spin dynamics in terms of localized moments, as is done in Eq. \eqref{eq:E_LSM}, the spin waves are treated adiabatically relative to the electron dynamics, which requires neglecting spin-flip excitations. In addition, when using \textit{linear} spin wave theory, also magnon-magnon interactions are neglected and Eq. \eqref{eq:H^munu_q} thus gives the adiabatic spectrum of noninteracting magnons. 

In the case of a single magnetic site per unit cell, Eq. \eqref{eq:H^munu_q} further simplifies to
\begin{align}
    \hbar \omega(\mathbf{q}) = \frac{g\mu_{\mathrm{B}}}{M}\left[J(\mathbf{0}) - J(\mathbf{q})\right]. \label{eq:single_site_energy}
\end{align}

\subsection{Curie temperatures}

Thermal excitation of magnon modes is the primary mechanism by which magnets disorder at finite temperatures $T$,
and the Curie temperature $T_{\mathrm{C}}$ of ferromagnets can be estimated from the magnon spectrum. With a single magnetic site in the unit cell and under a mean-field approximation \cite{Turek06_Heisenberg_model_for_itinerant_magnets}
\begin{align}
    k_{\mathrm{B}} T_{\mathrm{C}}^{\mathrm{MFA}} = \frac{M}{3g \mu_{\mathrm{B}}} \frac{1}{N_q} \sum_{\mathbf{q}} \hbar \omega(\mathbf{q}),  \label{eq:T_C^MFA}
\end{align}
where $k_{\mathrm{B}}$ is the Boltzmann constant and $N_q$ is the number of sampled $q$-points. If the localized spin fluctuations are instead treated at the RPA level, one finds \cite{Rusz05_RPA_for_T_c_of_collinear_magnets}
\begin{align}
    \frac{1}{k_{\mathrm{B}} T_{\mathrm{C}}^{\mathrm{RPA}}} = \frac{3g\mu_{\mathrm{B}}}{M} \frac{1}{N_q} \sum_{\mathbf{q}} \frac{1}{\hbar \omega(\mathbf{q})} \label{eq:T_C^RPA}.
\end{align}
In general, the RPA expression is expected to provide a better estimate for the Curie temperature and it can be shown that one always has $T_{\mathrm{C}}^{\mathrm{RPA}}<T_{\mathrm{C}}^{\mathrm{MFA}}$ \cite{Rusz05_RPA_for_T_c_of_collinear_magnets}.

\subsection{GPAW implementation \label{subsec:GPAW_implementation}}
We have implemented a new module for the computation of $\bar{J}^{ab}(\mathbf{q})$ in the electronic structure code GPAW \cite{Mortensen2005GPAW,Enkovaara2010GPAW_review} based on the plane wave representation \eqref{eq:J_breve_formula}. The module relies on existing functionality to calculate $\chi'^{+-}_{\mathrm{KS}}(\mathbf{q})$ as described in the linear-response time-dependent DFT (LR-TDDFT) implementation presented in Ref. \cite{Skovhus2021Published_version}. In LR-TDDFT, the full many-body susceptibility $\chi^{+-}(\mathbf{q}, \omega)$ is calculated within a given approximation for the time-dependent xc potential and the magnon frequencies are obtained as peaks in the dissipative part of $\chi^{+-}$. The main computational expense of the method is the calculation of the dynamic Kohn-Sham susceptibility $\chi^{+-}_{\mathrm{KS}}(\mathbf{q}, \omega)$ as a function of frequency. In the present approach, the adiabatic magnon spectrum is obtained directly from the reactive part of the {\it static} Kohn-Sham susceptibility and thus provides a significant computational simplification compared to LR-TDDFT. This comes at the expense that certain itinerant electron effects are neglected, and that e.g. the spectral broadening due to Landau damping cannot be described.

For the site-kernels $K^a(\mathbf{q})$, entering Eq. \eqref{eq:J_breve_formula}, we have implemented three possible geometrical shapes, as listed in App. \ref{appsec:site_kernel_formulas}.
The sphere is a natural choice for atom-centered magnetic sites, a cylinder may be useful for 2D systems and a parallelepiped is the generic shape of primitive unit cells. The freedom to choose shape and size of integration regions is a double-edged sword. On one hand it leads to ambiguity in the definition of magnetic sites; on the other it gives flexibililty to the overall construction of the spin-lattice model.

Once $B^{\mathrm{xc}}$, $K^a(\mathbf{q})$ and $\chi'^{+-}_{\mathrm{KS}}(\mathbf{q})$ have been computed for a given $\mathbf{q}$-point, we obtain $\bar{J}^{ab}(\mathbf{q})$ from Eq. \eqref{eq:J_breve_formula} and compute the magnon energies by diagonalizing Eq. \eqref{eq:H^munu_q} with $M_a=M$ fixed to the total magnetic moment per magnetic atom. Fixing $M_a$ to the total moment of the unit cell is only warranted for ferromagnets with equivalent sublattice sites. In the general case, one would instead fall back to the rigid spin approximation \eqref{eq:rigid_spin_approx} in order to define the local moments.

Finally, the Curie temperature is estimated from Eqs. \eqref{eq:T_C^MFA} and \eqref{eq:T_C^RPA} by sampling $\bar{J}^{ab}(\mathbf{q})$ throughout the entire Brillouin zone.

\section{Results and discussion \label{sec:Results_and_discussion}}

\subsection{Computational details \label{subsec:computational_details}}

In the following, we present computations of magnon energies as described in \cref{subsec:GPAW_implementation} for the elemental ferromagnets Fe, Co and Ni. We compute the LDA ground state in the PW92 parametrization \cite{Perdew1992} using experimental lattice parameters, a plane wave basis and 
a $\Gamma$-centered uniform $k$-point grid with $N_k$ points. We include all bands in fully and partially filled atomic shells plus a number of empty shell bands. The resulting magnetic moments are 2.16, 1.68, 1.65 and 0.66 for Fe(bcc), Co(fcc), Co(hcp) and Ni(fcc) respectively. Based on the LDA ground state, the reactive part of the static Kohn-Sham susceptibility $\chi'^{+-}_{\mathrm{KS}}$ is computed as well as the exchange-correlation magnetic field $B^{\mathrm{xc}}$.
The ground state is easily converged with respect to the plane wave basis, whereas the calculation of $\bar{J}^{ab}(\mathbf{q})$ in Eq. \eqref{eq:J_breve_formula} is truncated using an energy cutoff $E_{\mathrm{G}}$ such that 
only reciprocal lattice vectors with $\mathbf{G}^2/2 \leq E_{\mathrm{G}}$ are included.

We note that the susceptibility calculation requires $\mathbf{q}$ to be commensurate with the $k$-point grid of the underlying DFT ground state. This means that we can compute $\hbar \omega(\mathbf{q})$ only for $\mathbf{q}$ belonging to the $\Gamma$-centered $k$-point grid.

\subsection{Convergence tests: Fe, Co and Ni}

\begin{figure}[t]
    \centering
    \includegraphics[scale=0.9]{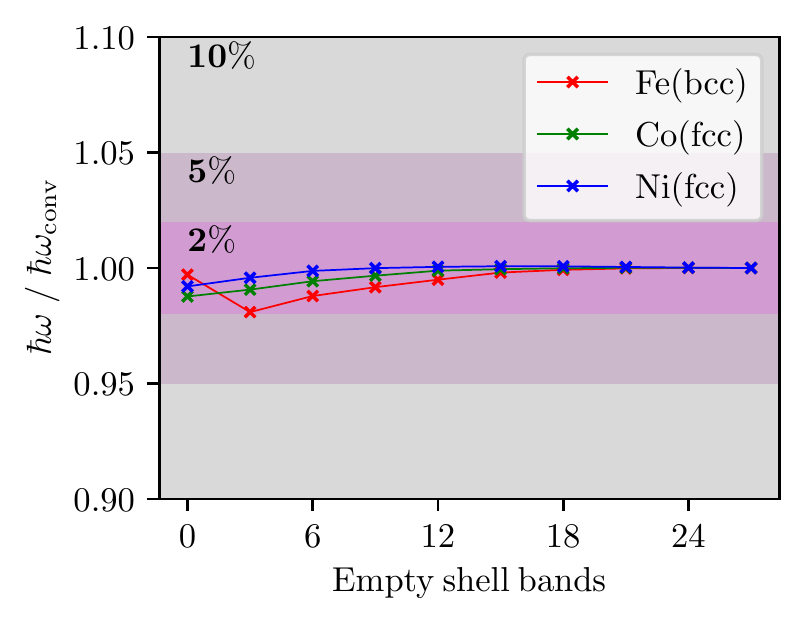}
    \includegraphics[scale=0.9]{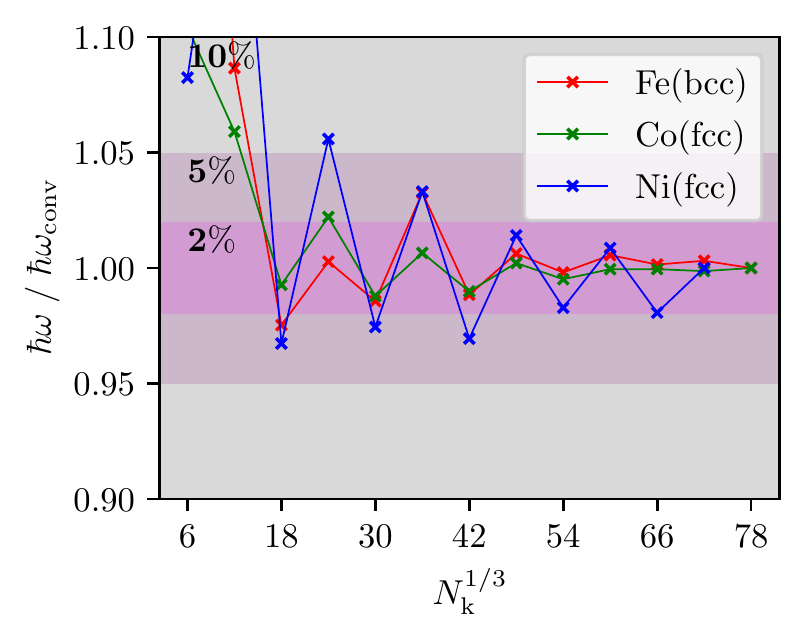}
    \includegraphics[scale=0.9]{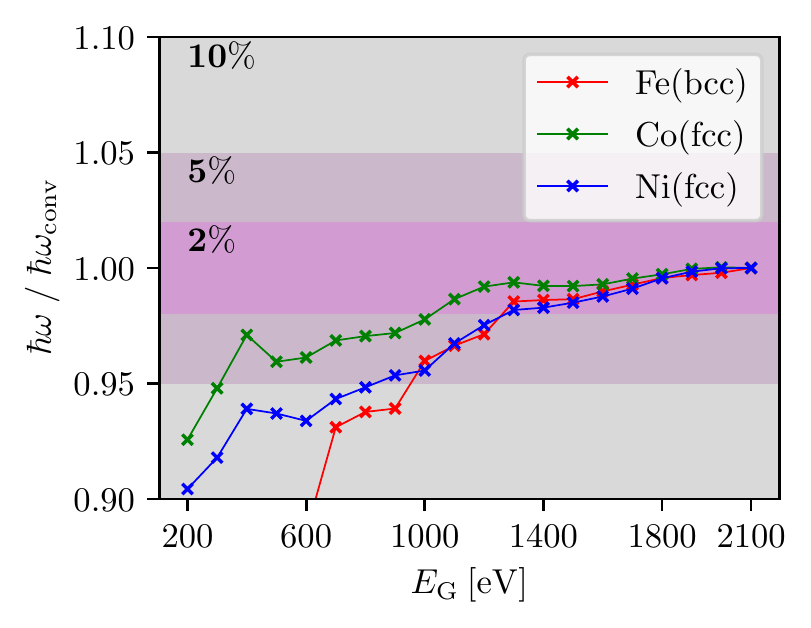}
    \caption{Convergence of magnon energies as a function of empty shell bands (top), $k$-points (middle) and plane wave cutoff (bottom). The magnon energies are calculated at high symmetry points ($\mathrm{X}$ for fcc and $\mathrm{H}$ for bcc) and normalized by the calculation performed at the highest level of completeness, $\hbar \omega_{\mathrm{conv}}$.
    Regions of $\leq$ 2, 5 and 10 \% deviation from $\hbar \omega_{\mathrm{conv}}$ are indicated by the background color.}
    \label{fig:convergence_study}
\end{figure}

The most critical parameters to converge are the number of empty shell bands, the number of $k$-points $N_k$ and the plane wave cutoff $E_{\mathrm{G}}$. In figure \ref{fig:convergence_study}, we present the magnon energies as a function of these parameters at the high-symmetry point $\mathrm{X}$ for fcc crystals and $\mathrm{H}$ for bcc crystals.
We use integration spheres with $r_{\mathrm{c}} = 1.3\:\text{Å}$.
In all three materials, the $N_k$ convergence fluctuates, falling within 5\% deviation at $N_k^{1/3}\sim 20$. The number of bands is already well converged even without empty shell bands, while $E_{\mathrm{G}} \sim 1000 \: \mathrm{eV}$ is required for a $\leq 5\%$ deviation. 
Typically, one would expect that a much more dense $k$-space sampling is required for metals in comparison to insulators, due to the presence of low-frequency Stoner pair excitations.

\subsection{Influence of the integration volume \label{subsec:Effect_of_integration_region}}

\begin{figure}[tb]
\centering
  \includegraphics[scale=0.9]{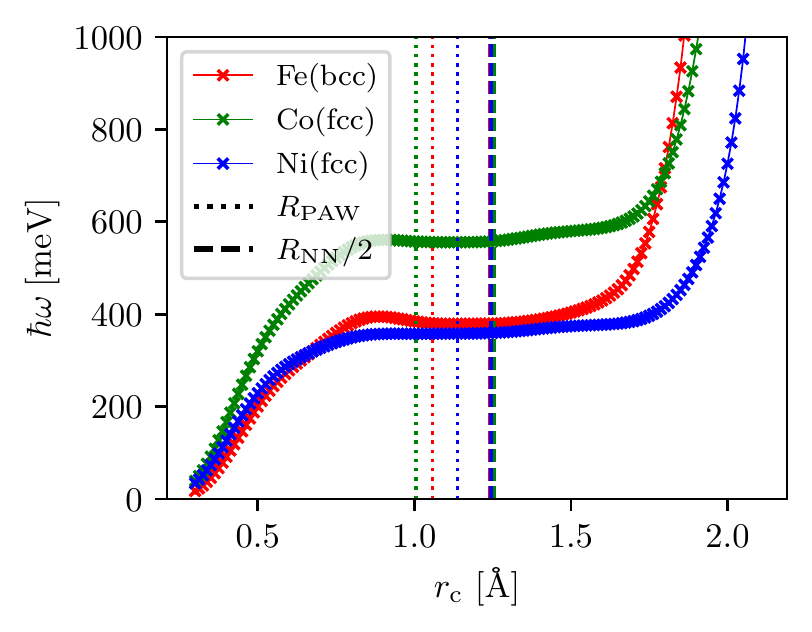}
\caption{Magnon energies at high symmetry points ($\mathrm{X}$ for fcc and $\mathrm{H}$ for bcc) for integration spheres of different radii. The radii of the $d$-orbital PAW sphere, $R_{\mathrm{PAW}}$ and half the nearest neighbour distance $R_{\mathrm{NN}}/2$ are shown as vertical lines.}\label{sfig:rc_conv}
\end{figure}

In figure \ref{sfig:rc_conv} we present magnon energies at the X and H high symmetry points of Fe(bcc), Co(fcc) and Ni(fcc) as a function of the radius $r_{\mathrm{c}}$ of the applied atom-centered spherical integration volume. As $r_{\mathrm{c}} \xrightarrow{} 0$, all energy contributions to the continuous model \eqref{eq:E_SM_cont} lie in the (neglected) interstitial region and 
$\hbar \omega \xrightarrow{} 0$.
Conversely, when there is a major overlap between the integration spheres of multiple atoms, the energy diverges. 
In between these two extremes, there is a plateau where $\hbar \omega$ is rather insensitive to $r_{\mathrm{c}}$. Looking at where the plateaus start, we conclude that the exchange contributions relevant for the spin wave dispersion can be effectively separated into localized spheres of radii $R^{\mathrm{Fe}}_{\mathrm{loc}} \approx 0.95\:\text{Å}$, $R^{\mathrm{Co}}_{\mathrm{loc}} \approx 0.85\:\text{Å}$ and $R^{\mathrm{Ni}}_{\mathrm{loc}} \approx 0.7\:\text{Å}$ respectively.
In all three materials, the nearest neighbour distance is roughly $R_{\mathrm{NN}} = 2.5\: \text{Å}$. 
We note that the energy begins to diverge around $R_{\mathrm{NN}} - R_{\mathrm{loc}}$, which is the point where the integration spheres start to overlap with the 3$d$-electrons of multiple atoms. The most natural choice for $r_{\mathrm{c}}$ in these cases is  $R_{\mathrm{NN}}/2$, which is the point where neighbouring integration spheres touch. From figure \ref{sfig:rc_conv} we see that $R_{\mathrm{NN}}/2$ falls more or less exactly at the middle of the plateau and the evaluated magnon energies are largely independent of the choice of $r_{\mathrm{c}}$ in the range $R_{\mathrm{NN}}/2\pm 0.2 $ \AA.

{\it A priori} it is not clear that a localized spin model like the Heisenberg model can provide a sufficient (or even a well defined) description of the magnetic properties in itinerant magnets - even if long range exchange interactions are included. The spin-polarized homogeneous electron gas, for example, does not allow for a unique partitioning of the spatial representation. For materials involving transition metal atoms, one might expect some degree of localization of the magnetic moments, since these are carried mostly by localized $d$-orbitals, but since Wannier functions are not exponentially localized in metals, it is not clear to what extent site-based modelling is applicable. The flat plateau in figure \ref{sfig:rc_conv} not only shows that the ambiguity in defining magnetic sites is not a problem in practice; it also implies that the {\it model} itself is well defined and well justified. The conclusion here is not necessarily transferable to other itinerant magnets, but the $r_{\mathrm{c}}$-dependence of magnon energies may in general be used as an analysis tool to assess the validity of localized spin models when calculating magnon energies. In this regard, it should be noted that it is computationally inexpensive to compute the magnon dispersion for a range of different cutoff radii $r_{\mathrm{c}}$, as the site-kernel \eqref{eq:site_kernels} is evaluated analytically.

\subsection{Magnon dispersion relations of Fe, Co and Ni}

\begin{figure*}[t]
    \centering
    \includegraphics[width=\linewidth]{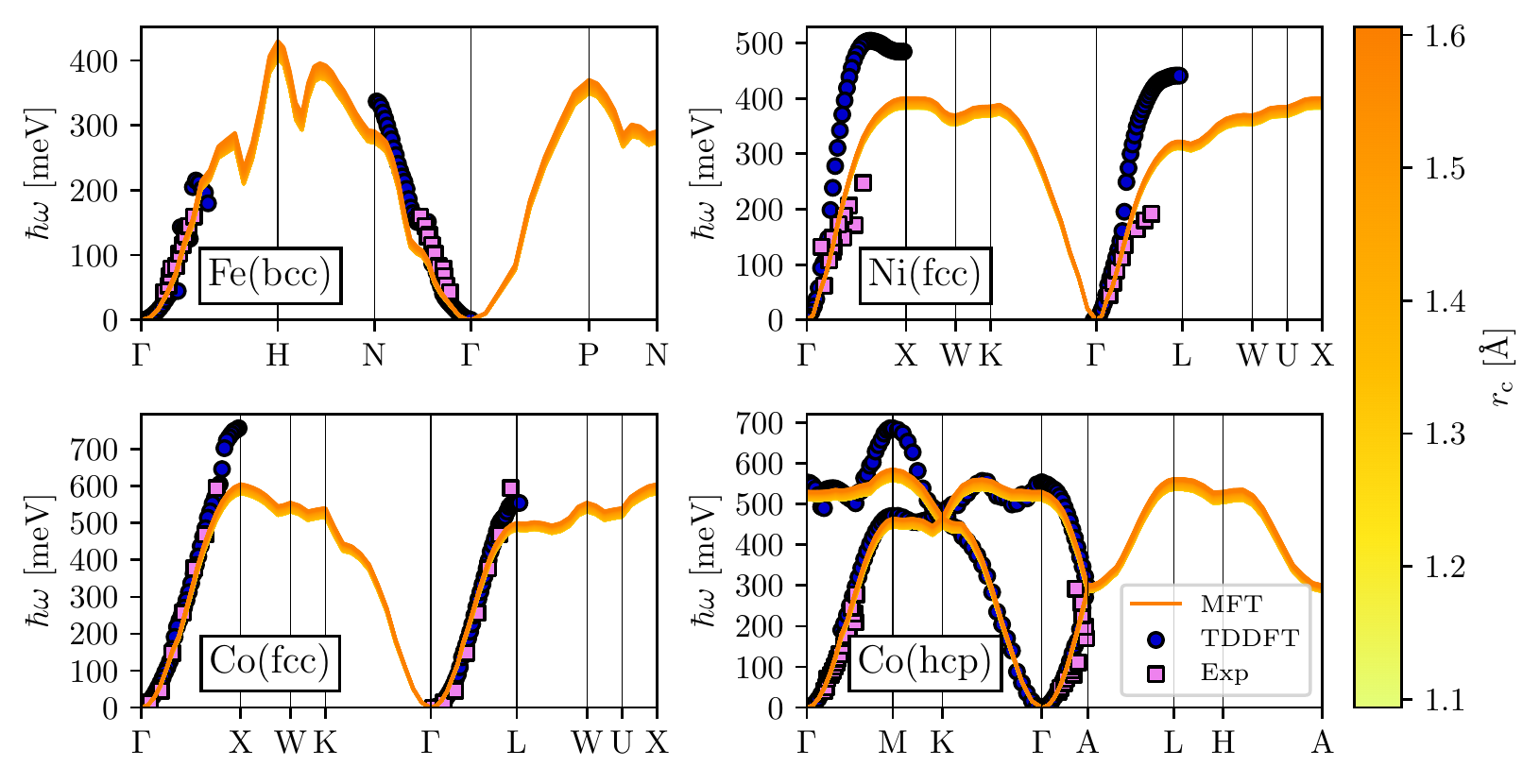}
    \caption{Adiabatic magnon dispersion relations (MFT) for Fe(bcc), Co(fcc), Co(hcp) and Ni(fcc), with comparison to TDDFT results from Ref. \cite{Skovhus2021Published_version} and experiments \cite{Loong1984,Mook1985,Balashov2009,Perring1995}.
    Computed with integration spheres of 129 different radii (color coded).}
    \label{fig:dispersion_relations}
\end{figure*}

In figure \ref{fig:dispersion_relations}, we present dispersion relations of elemental ferromagnets Fe, Co and Ni, computed with 
$E_{\mathrm{G}} = 1000\:\mathrm{eV}$ and 9 empty shell bands. We use $N_k^{1/3} = 30$ for Co(hcp) and $N_k^{1/3}=32$ for Co(fcc), Ni(fcc) and Fe(bcc), which includes all high-symmetry points of the lattices. Note that Co(hcp) has two atoms in the unit cell, why there are two magnon branches. Atom-centered spheres are used as integration volumes, with cutoff radii color coded.
We note that changing $r_{\mathrm{c}}$ in a span of $0.5\:\text{Å}$ has minimal influence on the dispersion shape and minor influence on numerical values, especially at low energies. This supports the conclusion of Sec. \ref{subsec:Effect_of_integration_region}, that the most relevant spin degrees of freedom are indeed localized in nature. 

The computed curves are in good agreement with previous calculations using either the frozen magnon approach \cite{Halilov1998_adiabatic_spin_dynamics_from_SDFT, VanSchilfgaarde1999} or a direct evaluation of the exchange parameters in real space \cite{VanSchilfgaarde1999,Pajda_2001_ab_init_exchange,Etz2015ASD_and_surface_magnons}. In Ref. \cite{Pajda_2001_ab_init_exchange} they do, however, find somewhat larger magnon bandwidths than we find in the present work. 

In Ref. \cite{Pajda_2001_ab_init_exchange} it was noted that the real space exchange constants decay extremely slowly as a function of distance between atoms, which makes the magnon stiffness (second derivative of dispersion at the $\Gamma$-point) difficult to converge. In the present work, this is reflected in the oscillating behaviour of magnon energies as a function of $k$-point sampling (see figure \ref{fig:convergence_study}). For Co(fcc) and Fe(bcc), we obtain a reasonable fit to an isotropic parabolic magnon dispersion $Dq^2$ in the long wavelength limit, when including all $q$-points on a $N_k^{1/3} = 36$ grid for which $q\leq 0.22\:\mathrm{\AA}^{-1}$. Using a cutoff radius corresponding to half the nearest neighbour distance, $R_{\mathrm{NN}}/2$, this results in magnon stiffnesses of $D=510\:\mathrm{meV}\,\mathrm{\AA}^2$ and $D=220\:\mathrm{meV}\,\mathrm{\AA}^2$ (rounded to nearest $10\:\mathrm{meV}\,\mathrm{\AA}^2$) for Co(fcc) and Fe(bcc) respectively. These results are in reasonable agreement with the TDDFT values of $D=490\:\mathrm{meV}\,\mathrm{\AA}^2$ and $D=250\:\mathrm{meV}\,\mathrm{\AA}^2$ estimated in Ref. \cite{Buczek2011}. For Ni(fcc), we are not able to produce a reliable fit of the magnon stiffness with a $k$-point sampling of $N_k^{1/3} = 36$, due to the difficulty of converging individual magnon energies, see figure \ref{fig:convergence_study}.

In the $q\rightarrow0$ limit TDDFT and MFT are expected to yield identical results, whereas the MFT method starts introducing systematic errors in the magnon dispersion at larger wave vectors \cite{Bruno03_renormalised_magnetic_force_theorem, Antropov2003}. In figure \ref{fig:dispersion_relations} we compare the MFT magnon dispersion to the dispersion extracted from LR-TDDFT within the same numerical implementation \cite{Skovhus2021Published_version} and show as well the experimental magnon energies inferred from inelastic scanning tunnelling spectroscopy (fcc-Co) and inelastic neutron scattering (the rest). Since the MFT method can be regarded as an adiabatic approximation to TDDFT \cite{Katsnelson_2004_Magnetic_susceptibility_LSDA} it is most illuminating to focus on the differences between TDDFT and MFT. In the case of bcc-Fe, it is important to distinguish between different directions in reciprocal space. Along the $\Gamma\rightarrow\mathrm{N}$ direction, the overall shape of the dispersion is in rather good agreement with TDDFT results \cite{Buczek2011,Singh2019, Rousseau2012,Cao2018,Skovhus2021Published_version}, and we find a magnon energy at $\mathrm{N}$ of roughly 285 meV, whereas TDDFT (in the ALDA) yields 335 meV when calculated using the same GPAW module as applied in the present work \cite{Skovhus2021Published_version}. 
Other calculations have reported values in the range of 280-365 meV \cite{Buczek2011,Singh2019, Rousseau2012,Cao2018}. Importantly, the plateau at $\sim0.5\,\mathbf{q}_\mathrm{N}$, which were also reported in Refs. \cite{Rousseau2012, Cao2018, Buczek2011,singh2018_story_of_magnetism} is well captured by MFT, albeit at a slightly lower energy. This feature originates from coupling to the Stoner continuum and results in complex spectral features in the dynamical susceptibility. Along the $\Gamma\rightarrow\mathrm{H}$ direction, the TDDFT spectrum is strongly influenced by Stoner pair excitations along most of the path. The magnon dispersion is in three consecutive instances subject to a spectral broadening followed by a jump/discontinuity (the last one leading to magnon frequencies in the range of 500-600 meV) at points where the magnon mode crosses a stripe-like feature in the noninteracting Stoner spectrum. After the third jump, there is a significant increase of the linewidth, and around $\sim0.6\,\mathbf{q}_{\mathrm{H}}$, the magnon mode is completely suppressed by the Stoner pair excitations. In general, the dynamical susceptibility entails a significant amount of information, which is discarded in the adiabatic approximation. In particular, the magnon energy can only be rigorously  identified from TDDFT when the spectral function has a Lorentzian lineshape, but if the linewidth becomes large a strong coupling to the Stoner continuum is implied and the adiabatic approximation becomes questionable. For the itinerant magnets studied here, the 
linewidth is on the order of hundred(s) of meV at the Brillouin zone boundary. Moreover, at the $\mathrm{H}$-point in bcc-Fe the lineshape is far from Lorentzian and the extraction of a magnon energy from the spectral function thus becomes dubious. It is therefore not clear if the $\Gamma\rightarrow\mathrm{H}$ adiabatic magnon spectrum obtained from MFT has any physical significance (except in the limit $\mathbf{q}\rightarrow0$). It is, however, remarkable that the continuous MFT dispersion still manages to roughly predict the three disconnected magnon energies along this path (but not the one at 500-600 meV closest to the $\mathrm{H}$-point).

In the case of Ni, we find a magnon bandwith of $\sim390$ meV, whereas the LR-TDDFT implementation in GPAW yields $505$ meV \cite{Skovhus2021Published_version} and others have reported TDDFT values in the range of 460-555 meV \cite{Buczek2011,Singh2019, Rousseau2012,Cao2018}.
For Ni, the TDDFT magnon spectrum exhibits a well defined magnon peak position throughout the entire BZ, although the magnon is strongly damped on the BZ boundary. Thus, the comparison between MFT and TDDFT results should be meaningful 
and the fact that the MFT result is much closer to the experimental value, which is on the order of 250 meV, is coincidental. Rather, the discrepancy between MFT and TDDFT results is likely due to the low magnetic moment (low exchange splitting) in Ni, which increases the importance of longitudinal spin fluctuations and the systematic errors entailed by the adiabatic approximation \cite{Sponza2017,Grotheer2001}. The poor agreement between experiments and the ALDA dispersion is also related to this point, since it can be attributed to the overestimation of the exchange splitting in LDA \cite{Sasoglu2010}, which is not properly corrected for in the dynamic susceptibility at the ALDA level. Once again, it is noteworthy that complex spectral features of itinerant nature (Stoner stripes and the associated magnon branch splitting) observed in the TDDFT and MBPT spectra close to the $\Gamma$-point \cite{Karlsson2000,Sasoglu2010,Buczek2011,Friedrich2020,Skovhus2021Published_version} emerge as kinks (Kohn anomalies) in the MFT magnon dispersion, which will always be continuous when solving the Heisenberg model within linear spin wave theory.

For hcp Co, the present MFT calculations agree very well with previous TDDFT results for the acoustic branch and the degeneracy point at $\mathrm{K}$ is located at $\sim455$ meV in MFT and 475 meV in TDDFT. The optical branch shows good agreement as well, although MFT predict somewhat lower magnon energies around the maximum at $\mathrm{M}$, which is located at $\sim575$ meV in MFT and at 680 meV in TDDFT \cite{Buczek2011,Skovhus2021Published_version}. Similarly, for fcc Co we find that the MFT dispersion only starts to deviate from the corresponding TDDFT results when approaching the Brillouin zone boundary, yielding a maximum magnon energy of $\sim595$ meV at the $\mathrm{X}$-point whereas TDDFT yields a bandwidth of 755 meV \cite{Skovhus2021Published_version}. 

In general, the MFT magnon dispersion relations calculated in the LDA should be viewed as approximations to the ALDA TDDFT dispersion. It has previously been established that adiabatic localized spin model descriptions, such as the MFT formalism applied here, are formally correct only in the long wavelength limit \cite{Edwards1985,Katsnelson2000,Muniz2002}. In figure \ref{fig:dispersion_relations}, we reproduce this finding in practise with deviations from the TDDFT dispersion arising at finite wave vectors $\mathbf{q}$, as it has also been confirmed by previous comparisons between MFT and TDDFT \cite{Buczek2011}. Apart from the adiabatic approximation itself, which is generally expected to break down as the magnon mode enters the Stoner continuum, the standard MFT formalism also includes systematic errors due to the fact that the rotated states treated perturbatively do not include any constraining fields. This can be traced back to Eq. \eqref{eq:E_SM_cont}, where the correspondence between the spin model and DFT requires a constrained functional $E[\mathbf{u}]$. Inclusion of such fields leads to a renormalized magnetic force theorem \cite{Bruno03_renormalised_magnetic_force_theorem,Antropov2003}, typically increasing the predicted values of $J$ and giving rise to larger magnon energies in ferromagnets. Furthermore, the constraining field renormalization itself appears to be connected to how the adiabatic limit is defined \cite{Katsnelson_2004_Magnetic_susceptibility_LSDA} and might be closely related to the breakdown of the adiabatic approximation discussed above. The exact connection between the renormalization and nonadiabatic effects in the dispersion is not fully clear in our present understanding. In the future, it would be highly interesting to perform a systematic comparison of both standard and renormalized MFT to TDDFT calculations in order to clarify the connection further.

\subsection{Critical temperatures}
\begin{table}[ht]
\begin{tabular}{l|l|l|l|l|l}
\toprule
& $T_\mathrm{C}^\mathrm{MFA}$ & $T_\mathrm{C}^\mathrm{RPA}$ & $T_\mathrm{C}^\text{Halilov}$\cite{Halilov1998_adiabatic_spin_dynamics_from_SDFT} & $T_\mathrm{C}^\mathrm{DMFT}$\cite{Liechtenstein2001FiniteTemperatureMagnetismDMFT} & $T_\mathrm{C}^\mathrm{exp}$\cite{kittel1996introduction} \\ \hline
Fe(bcc)              & 1020        & 676       & 1037   & 1900              & 1043       \\ 
Co(fcc)               & 1353       & 1062        & 1250  & -                & 1388       \\ 
Ni(fcc)               & 375        & 323        & 430   & 700                & 627        \\ 
    \bottomrule
\end{tabular}
\caption{Curie temperatures in Kelvin, $T_\mathrm{C}^\mathrm{MFA}$ and $T_\mathrm{C}^\mathrm{RPA}$, calculated using the MFT magnon dispersion and compared to experiment $T_\mathrm{C}^\mathrm{exp}$ and previous theoretical results $T_\mathrm{C}^{\text{Halilov}}$ and $T_\mathrm{C}^\mathrm{DMFT}$.}
\label{tab:critical_temperatures}
\end{table}

In table \ref{tab:critical_temperatures}, we compare computed critical temperatures with experiments, the results of Halilov et\ al. based on frozen magnon calculations \cite{Halilov1998_adiabatic_spin_dynamics_from_SDFT} and the results of Liechtenstein et\ al. \cite{Liechtenstein2001FiniteTemperatureMagnetismDMFT}, who calculated the finite-temperature magnetization self-consistently in the framework of Dynamical Mean-Field Theory. The Curie temperatures are calculated by integrating Eqs. \eqref{eq:T_C^MFA} and \eqref{eq:T_C^RPA} on the full $N_q^{1/3} = 36$ $q$-point grid commensurate with the underlying $N_k^{1/3} = 36$ $\Gamma$-centered Monkhorst-Pack grid of the ground state.
We use a spherical site cutoff radius of $r_\mathrm{c}=R_{\mathrm{NN}}/2$. 
The computation was greatly simplified by the high degree of symmetry in fcc and bcc lattices, as we only needed to compute $q$-points in the irreducible part of the Brillouin zones.

The agreement with experiments is seen to be rather poor for $T_\mathrm{C}^\mathrm{RPA}$, whereas the mean-field approximation yields surprisingly accurate Curie temperatures, except in the case of Ni. The results of Halilov et al.\ \cite{Halilov1998_adiabatic_spin_dynamics_from_SDFT} are evaluated from the mean field approximation and are fairly consistent with our $T_\mathrm{C}^\mathrm{MFA}$ values. In principle, $T_\mathrm{C}^\mathrm{RPA}$ should be more accurate, but it is far from clear how well the approximation performs in general. To this end, we note that MFT tends to underestimate magnon dispersion relations (compared to TDDFT), which results in an {\it underestimation} of critical temperatures.
Although RPA is routinely used to predict critical temperatures, it is only expected to be a good approximation at low temperatures. Moreover, it is derived from a model of strictly localized spins and it is not obvious how nonadiabatic effects influence the predictions of RPA - even if we are able to accurately calculate the magnon dispersion.

\section{Summary \label{sec:Conclusion}}
We have presented a derivation of the microscopic exchange tensor based on the magnetic force theorem. The method was in fact derived 35 years ago for the isotropic case using multiple scattering theory and has subsequently been generalized to include various specific spin-orbit effects such as Dzyaloshinskii-Moriya interactions. However, a derivation of the full exchange tensor without additional assumptions was, to our knowledge, missing in literature. In the present work, we have taken a general noncollinear system (including spin-orbit interactions) and derived a basis independent relation between the microscopic exchange tensor and the transverse magnetic susceptibility of the Kohn-Sham system. The derivation is based on second order time-independent perturbation theory and provides, in our opinion, a much simpler approach than the original formulation.

The method has been implemented (without spin-orbit coupling) in the GPAW open-source code using a plane wave basis, which allows for systematic convergence with respect to basis set and unoccupied states. Since several electronic structure codes can perform calculations of susceptibilities in a plane wave basis, we believe that the present formulation may be highly useful and allow for a seamless implementation into existing DFT codes. In addition, the microscopic exchange tensor allows for complete freedom in the definition of magnetic sites relevant to a given problem. We find it a virtue that the magnetic sites can be defined based solely on the magnetization density (which is a physical quantity) rather than localized orbitals derived from Kohn-Sham states.

For the elemental ferromagnets Fe, Co and Ni, we applied the method to calculate the adiabatic spin wave spectrum in the local density approximation and find good agreement with calculations from the literature. Importantly, we showed that the spin wave energies are largely insensitive to the choice of magnetic sites, which to some extent validates the application of the Heisenberg model despite the itinerant nature of the materials. The results were compared to spectra obtained with TDDFT (using the ALDA) and the deviations of the results obtained with the two methods were discussed in the context of the adiabatic approximation. The formulation of exchange constants in terms of the magnetic susceptibility renders the extension of standard MFT to the renormalized MFT of Bruno \cite{Bruno03_renormalised_magnetic_force_theorem} straightforward, but we leave a detailed comparison with that theory for future work.

\section{Acknowledgement}
The authors acknowledge support from the Villum foundation Grant No. 00029378

\onecolumngrid

\appendix

\section{The four-component susceptibility tensor\label{appsec:Standard_and_rotated_susceptibility}}

The four-component susceptibility tensor $\chi^{\mu\nu}(\mathbf{r}, \mathbf{r}', \omega)$ describes the linear response in electron density and spin-polarization when a material is perturbed by external electric and magnetic fields. In the context of mapping the DFT energy functional to a classical spin model, the effect of orbital and lattice degrees of freedom are neglected and the relevant response properties are encoded entirely by $\chi^{\mu\nu}$. In general, the susceptibility is a dynamic (retarded) quantity, which can be split into reactive and dissipative parts $\chi^{\mu\nu}(\mathbf{r}, \mathbf{r}', \omega)=\chi'^{\mu\nu}(\mathbf{r}, \mathbf{r}', \omega)+i\chi''^{\mu\nu}(\mathbf{r}, \mathbf{r}', \omega)$. The dissipative part is composed of $\delta$-function peaks at frequencies corresponding to eigenstate transitions in the material and is related to the reactive part by a Kramers-Kronig relation \cite{Skovhus2021Published_version}. It turns out that in order to calculate the exchange tensor in the framework of MFT, only the static limit ($\omega\rightarrow 0$) of the reactive part of the noninteracting Kohn-Sham susceptibility is needed. This quantity can be computed based solely on Kohn-Sham quantities extracted from a DFT ground state calculation \cite{Skovhus2021Published_version},
\begin{align}
    \chi_{\mathrm{KS}}'^{\mu\nu}(\mathbf{r}, \mathbf{r}') =  \sum_n\sum_{m\neq n} \frac{f_n - f_m}{\varepsilon_n - \varepsilon_m} \sum_{s, t} \sum_{s', t'} \sigma^{\mu}_{s t} \sigma^{\nu}_{s' t'} \psi_{n,s}^*(\mathbf{r}) \psi_{m,t}(\mathbf{r}) \psi^*_{m,s'}(\mathbf{r}')\psi_{n,t'}(\mathbf{r}') ,\label{eq:chi_general}
\end{align}
where $s,t, s', t' \in \{\uparrow, \downarrow\}$ and $\psi_{n,s}$ denotes the $s$ spin component of the $n$'th Kohn-Sham (spinorial) eigenstate with single-particle energy $\varepsilon_n$ and occupation factor $f_n$. In this work, we make use of two different basis representations for the spin degrees of freedom. In the standard cartesian representation, $\mu, \nu \in \{0,x,y,z\}$, $\sigma^\mu$ simply denotes the Pauli matrices augmented by the identity:
\begin{align*}
    \sigma^0 = \begin{pmatrix}1 & 0 \\ 0 & 1\end{pmatrix},\qquad
    \sigma^x = \begin{pmatrix}0 & 1 \\ 1 & 0\end{pmatrix},\qquad 
    \sigma^y = \begin{pmatrix}0 & -i \\ i & 0\end{pmatrix},\qquad
    \sigma^z = \begin{pmatrix}1 & 0 \\ 0 & -1\end{pmatrix}.
\end{align*}
The $\chi^{00}$ component characterizes the dielectric response, the cartesian components $\chi^{\alpha\beta}$ with $\alpha,\beta\in\{x,y,z\}$ govern the magnetic response, while combinations $\chi^{0\alpha}$ and $\chi^{\alpha 0}$ yield the linear response in the dielectric component to magnetic components of the external field and vice-versa. It is, however, often more convenient to use a spin-flip representation of the susceptibility \eqref{eq:chi_general}, where the cartesian representation $\mu,\nu$ is replaced with spin components $a,b \in \{\uparrow, \downarrow, +, -\}$ where
\begin{align*}
    \sigma^{\uparrow} = \begin{pmatrix}1 & 0 \\ 0 & 0\end{pmatrix},\qquad \sigma^{\downarrow} = \begin{pmatrix}0 & 0 \\ 0 & 1\end{pmatrix},\qquad \sigma^{+} = \begin{pmatrix}0 & 1 \\ 0 & 0\end{pmatrix},\qquad \sigma^- = \begin{pmatrix}0 & 0 \\ 1 & 0\end{pmatrix}.
\end{align*}
It is straightforward to write the susceptibility in the cartesian representation as linear combinations of the spin-flip susceptibilities. For example, $\chi^{xx}=\chi^{+-}+\chi^{-+}+\chi^{--}+\chi^{++}$ and $\chi^{xy}=i\chi^{+-}-i\chi^{-+}+i\chi^{--}-i\chi^{++}$, which are valid for the full dynamic many-body susceptibility as well as the Kohn-Sham susceptibility \cite{Skovhus2021Published_version}.

The spin-flip representation becomes particularly useful for collinear systems without spin-orbit coupling. In this case, the total spin projection along the $z$-direction $S^z$ can be taken as a good quantum number, why $\chi^{--}=\chi^{++}=0$ \cite{Skovhus2021Published_version}. Furthermore, one may choose the Kohn-Sham orbitals to be real, why $\chi_{\mathrm{KS}}'^{+-}(\mathbf{r},\mathbf{r}')=\chi_{\mathrm{KS}}'^{-+}(\mathbf{r},\mathbf{r}')$. Combining these expressions then yields a significant simplification of the transverse components, $\chi_{\mathrm{KS}}'^{xx}(\mathbf{r},\mathbf{r}')=\chi_{\mathrm{KS}}'^{yy}(\mathbf{r},\mathbf{r}')=2\chi_{\mathrm{KS}}'^{+-}(\mathbf{r},\mathbf{r}')$ and $\chi_{\mathrm{KS}}'^{xy}(\mathbf{r},\mathbf{r}')=\chi_{\mathrm{KS}}'^{yx}(\mathbf{r},\mathbf{r}')=0$.
Similarly, the transverse $x,y$ and longitudinal $0,z$ components do not couple whenever $S^z$ is a good quantum number (e.g. $\chi^{0y}=\chi^{y0}=0$) \cite{Skovhus2021Published_version}, which ultimately leads to the conclusion that $\mathrm{Im}\chi_{\mathrm{KS}}'^{\mu\nu}(\mathbf{r},\mathbf{r}')=\mathrm{Im}\chi_{\mathrm{KS}}'^{ab}(\mathbf{r},\mathbf{r}')=0$.

\section{Evaluation of second order energy contributions in the LDA\label{appsec:Delta_E_DFT}}
Here we evaluate explicitly the leading order energy cost $\Delta E_{\mathrm{DFT}}^{(2)}$ of performing a rotation of the local magnetization direction away from the ground state (see Eq. \eqref{eq:Delta_E_DFT}) in terms of Kohn-Sham quantities.

\subsection{Longitudinal contribution\label{appsec:first_term_of_Delta_E_DFT}}

The first term of $\Delta E_{\mathrm{DFT}}^{(2)}$ in Eq. \eqref{eq:Delta_E_DFT} gives the change in energy due to the reduced local moment along the local magnetization direction. The matrix elements entering this longitudinal contribution can be written out explicitly as
\begin{align*}
    \langle\psi_n| \delta^2 H_{\mathrm{KS}} |\psi_n\rangle = -\frac{1}{2} \int \mathrm{d}\mathbf{r}\, \theta(\mathbf{r})^2 B_{\mathrm{xc}}(\mathbf{r}) \psi_n^{\dag}(\mathbf{r}) \Tilde{\sigma}_z(\mathbf{r}) \psi_n(\mathbf{r}) = -\frac{1}{2} \int \mathrm{d}\mathbf{r}\, \theta(\mathbf{r})^2 B_{\mathrm{xc}}(\mathbf{r}) \left(|\Tilde{\psi}_{n,\uparrow}(\mathbf{r})|^2 - |\Tilde{\psi}_{n,\downarrow}(\mathbf{r})|^2\right),
\end{align*}
where $\Tilde{\psi}_n(\mathbf{r})=\mathrm{U}_{1/2}(\mathbf{r})\psi_n(\mathbf{r})$, see Eq. \eqref{eq:locally rotated pauli vector}.
In the local frame of reference, the magnetization is aligned along the $z$-axis and the magnitude of the magnetization can thus always be written as
\begin{align*}
    m_0(\mathbf{r}) = \sum_n^{\mathrm{occ}} \left(|\Tilde{\psi}_{n,\uparrow}(\mathbf{r})|^2 - |\Tilde{\psi}_{n,\downarrow}(\mathbf{r})|^2\right),
\end{align*}
such that
\begin{align}
    \Delta E_{\mathrm{DFT}}^{(2,\parallel)} = \sum_n^{\mathrm{occ}} \langle\psi_n| \delta^2 H_{\mathrm{KS}} |\psi_n\rangle = -\frac{1}{2} \int \mathrm{d}\mathbf{r}\, \theta(\mathbf{r})^2 B_{\mathrm{xc}}(\mathbf{r}) m_0(\mathbf{r}). \label{eq:Delta_E_DFT_first_term}
\end{align}
Comparing this result to the equivalent longitudinal term in the classical spin model \eqref{eq:Delta_E^2_SM_continuous}, we can identify $\Tilde{J}^{zz}(\mathbf{r}, \mathbf{r}')$ according to Eq. \eqref{eq:Jzz_from_magnetisation}.

\subsection{Transverse contribution\label{appsec:second_term_of_Delta_E_DFT}}

The transverse contribution to $\Delta E_{\mathrm{DFT}}^{(2)}$, i.e. the second term in Eq. \eqref{eq:Delta_E_DFT}, governs the change in energy from magnetization components perpendicular to the local ground state magnetization as induced by the rotation. As spin waves are transverse modes of excitations, this term contains the essential physics to describe the spin wave dispersion. Written out explicitly, the transverse contribution is given by:
\begin{align*}
    \Delta E_{\mathrm{DFT}}^{(2,\perp)}
    = \sum_n^{\mathrm{occ}} \sum_{m\neq n} \frac{\big|\langle\psi_m| B^{\mathrm{xc}} \delta \Tilde{\mathbf{u}} \boldsymbol{\cdot} \Tilde{\sigmaVec} |\psi_n\rangle \big|^2}{\varepsilon_n - \varepsilon_m}.
\end{align*}
Rewriting the sum over occupied states in terms of the ground state occupation numbers, $\sum_n^{\mathrm{occ}} (...) = \sum_n f_n (...)$, gives
\begin{align}
   \Delta E_{\mathrm{DFT}}^{(2,\perp)} &= \sum_n \sum_{m\neq n} \frac{f_n}{\varepsilon_n - \varepsilon_m} \iint \mathrm{d}\mathbf{r} \mathrm{d}\mathbf{r}'\, B_{\mathrm{xc}}(\mathbf{r}) B_{\mathrm{xc}}(\mathbf{r}') \left.\left[\psi_m^{\dag}(\delta \Tilde{\mathbf{u}} \boldsymbol{\cdot} \Tilde{\sigmaVec}) \psi_n \right]\right|_{\mathbf{r}} \left.\left[\psi_n^{\dag}(\delta \Tilde{\mathbf{u}} \boldsymbol{\cdot} \Tilde{\sigmaVec}) \psi_m \right]\right|_{\mathbf{r}'}\notag\\
    &= \frac{1}{2}\sum_n\sum_{m\neq n} \frac{f_n-f_m}{\varepsilon_n - \varepsilon_m} \iint \mathrm{d}\mathbf{r} \mathrm{d}\mathbf{r}'\, B_{\mathrm{xc}}(\mathbf{r}) B_{\mathrm{xc}}(\mathbf{r}') \left.\left[\psi_m^{\dag}(\delta \Tilde{\mathbf{u}} \boldsymbol{\cdot} \Tilde{\sigmaVec}) \psi_n \right]\right|_{\mathbf{r}} \left.\left[\psi_n^{\dag}(\delta \Tilde{\mathbf{u}} \boldsymbol{\cdot} \Tilde{\sigmaVec}) \psi_m \right]\right|_{\mathbf{r}'}\notag\\
    &= \frac{1}{2} \iint \mathrm{d}\mathbf{r} \mathrm{d}\mathbf{r}'\, B_{\mathrm{xc}}(\mathbf{r})\delta \Tilde{\mathbf{u}}^T(\mathbf{r})\tilde\chi'_\mathrm{KS}(\mathbf{r},\mathbf{r}')\delta \Tilde{\mathbf{u}}(\mathbf{r}') B_{\mathrm{xc}}(\mathbf{r}'),
    \label{eq:E2_transverse_final}
\end{align}
where it was used that the locally rotated Pauli vector $\Tilde{\sigmaVec}(\mathbf{r})$ is Hermitian, matrix multiplication is implied in the last equality and $\Tilde{\chi}'_\mathrm{KS}(\mathbf{r},\mathbf{r}')$ is the cartesian part of the susceptibility tensor in the local frame of reference. Specifically,
\begin{align}
    \Tilde\chi_{\mathrm{KS}}'^{\mu\nu}(\mathbf{r}, \mathbf{r}') =  \sum_n \sum_{m\neq n} \frac{f_n - f_m}{\varepsilon_n - \varepsilon_m} \sum_{s, t} \sum_{s', t'} \Tilde{\sigma}^{\mu}_{s t}(\mathbf{r}) \Tilde\sigma^{\nu}_{s' t'}(\mathbf{r}') \psi_{n,s}^*(\mathbf{r}) \psi_{m,t}(\mathbf{r}) \psi^*_{m,s'}(\mathbf{r}')\psi_{n,t'}(\mathbf{r}'), \label{eq:chi_rotated}
\end{align}
which can be obtained from the laboratory susceptibility \eqref{eq:chi_general} by taking $\sigma^{\mu}\rightarrow\tilde\sigma^{\mu}(\mathbf{r})=\mathrm{U}^\dag_{1/2}(\mathbf{r})\sigma^{\mu}\mathrm{U}_{1/2}(\mathbf{r})$.
Using once more that $\tilde\sigma^{\mu}(\mathbf{r})^\dagger=\tilde\sigma^{\mu}(\mathbf{r})$, it is straightforward to show that $\Tilde{\chi}'^\dag_\mathrm{KS}(\mathbf{r}',\mathbf{r})=\Tilde{\chi}'_\mathrm{KS}(\mathbf{r},\mathbf{r}')$, from which it is clear that only the real part of $\Tilde\chi'_\mathrm{KS}(\mathbf{r}',\mathbf{r})$ contributes to the energy difference:
\begin{equation}
    \Delta E_{\mathrm{DFT}}^{(2,\perp)} = \frac{1}{2} \iint \mathrm{d}\mathbf{r} \mathrm{d}\mathbf{r}'\, \delta \Tilde{\mathbf{u}}^T(\mathbf{r}) B_{\mathrm{xc}}(\mathbf{r}) \mathrm{Re}\{\tilde\chi'_\mathrm{KS}(\mathbf{r},\mathbf{r}')\}B_{\mathrm{xc}}(\mathbf{r}')\delta \Tilde{\mathbf{u}}(\mathbf{r}').
\end{equation}
By comparison with Eq. \eqref{eq:Delta_E^2_SM_continuous}, we can then immediately identify the transverse components of $\mathrm{J}(\mathbf{r},\mathbf{r}')$ according to Eq. \eqref{eq:J_cartesian_tilde}.

\section{\label{appsec:Relation_to_Greens_function_results}Relation to Green's function results}
The MFT linear response formula for the magnetic exchange tensor is most commonly formulated in terms of Green's functions. Here it is shown explicitly that the expressions are equivalent in the collinear case without spin-orbit coupling.
The Kohn-Sham Greens function can be written in the Lehmann representation as
\begin{align}
    G^{s}_\mathrm{KS}(\mathbf{r}, \mathbf{r}', \omega + i\eta) = \sum_n \frac{\psi_{ns}(\mathbf{r})\psi_{ns}^*(\mathbf{r}')}{\hbar\omega - \varepsilon_{ns} + i \hbar\eta} 
    \label{eq:Greens_function}
\end{align}
with $\eta>0$. 
Using partial fraction decomposition, we obtain
\begin{align}
    G^{\uparrow}_\mathrm{KS}(\mathbf{r}, \mathbf{r}', \omega+i\eta)G^\downarrow_\mathrm{KS}(\mathbf{r}', \mathbf{r}, \omega+i\eta) \simeq \sum_{n,m} \frac{\psi_{m\uparrow}(\mathbf{r})\psi_{m\uparrow}^*(\mathbf{r}')\psi_{n\downarrow}(\mathbf{r}')\psi_{n\downarrow}^*(\mathbf{r})}{\varepsilon_{n\downarrow} - \varepsilon_{m\uparrow}} \left[\frac{1}{\omega - \varepsilon_{n\downarrow} + i \hbar\eta} - \frac{1}{\omega - \varepsilon_{m\uparrow} + i\hbar\eta}\right]
    \label{eq:partial_decomposition}
\end{align}
up to an extra set of terms taking care of coincidental degeneracies, i.e.\ when $\varepsilon_{n\downarrow}=\varepsilon_{m\uparrow}$. Using that
$\lim_{\eta\rightarrow 0^+}\frac{1}{x+i\hbar \eta} = \frac{\mathcal{P}}{x} - i \pi \delta(x)$ and taking the Kohn-Sham orbitals as real, one finds that
\begin{align}
    -\frac{\hbar}{\pi}
    \lim_{\eta\rightarrow 0^+}
    \int_{-\infty}^{\varepsilon_{\mathrm{F}}/\hbar} \mathrm{d}\omega\, \mathrm{Im} G_\mathrm{KS}^{\uparrow}(\mathbf{r}, \mathbf{r}', \omega+i\eta) G_\mathrm{KS}^{\downarrow}(\mathbf{r}', \mathbf{r}, \omega+i\eta) &= \sum_{n,m} \frac{\psi_{n\downarrow}^*(\mathbf{r})\psi_{m\uparrow}(\mathbf{r})\psi_{m\uparrow}^*(\mathbf{r}')\psi_{n\downarrow}(\mathbf{r}')}{\varepsilon_{n\downarrow} - \varepsilon_{m\uparrow}} \left[ f_{n\downarrow} - f_{m\uparrow}\right] \notag \\
    &= \chi'^{
    -+}_{\mathrm{KS}}(\mathbf{r}, \mathbf{r}'),
    \label{eq:susceptibility_greens_fnc_relation}
\end{align}
where $\varepsilon_{\mathrm{F}}$ denotes the Fermi energy. Using that $\chi_{\mathrm{KS}}'^{+-}(\mathbf{r},\mathbf{r}')=\chi_{\mathrm{KS}}'^{-+}(\mathbf{r},\mathbf{r}')$ and inserting into Eq. \eqref{eq:exchange_in_ferromagnet} yields the Green's function MFT expression 
\begin{align*}
    J(\mathbf{r}, \mathbf{r}') = \frac{2\hbar}{\pi} \mathrm{Im} \int_{-\infty}^{\varepsilon_{\mathrm{F}}/\hbar} \mathrm{d}\omega\, B^{\mathrm{xc}}(\mathbf{r}) G_\mathrm{KS}^{\uparrow}(\mathbf{r}, \mathbf{r}', \omega) B^{\mathrm{xc}}(\mathbf{r}') G_\mathrm{KS}^{\downarrow}(\mathbf{r}', \mathbf{r}, \omega),
\end{align*}
see e.g. Ref. \cite{Bruno03_renormalised_magnetic_force_theorem}. It should be noted that prefactors may vary due to differences in the definition of the Heisenberg exchange parameters. Furthermore, we note that the expression \eqref{eq:partial_decomposition} becomes an equality for finite systems where $\varepsilon_{n\downarrow}$ and $\varepsilon_{m\uparrow}$ are nondegenerate. In the strict thermodynamic limit of periodic solids, where the eigenenergy spectrum is continuous, one obtains an additional sum involving terms with $\varepsilon_{n\downarrow}=\varepsilon_{m\uparrow}$. It is, however, straightforward to verify that such terms will not contribute to the frequency integrals above, so that the Eq. \eqref{eq:susceptibility_greens_fnc_relation} remains true also in this limit. 

\section{\label{appsec:site_kernel_formulas} Analytical site-kernels}\label{appsec:site_kernels}
In this section we provide closed form analytical solutions of the site kernel integrals in Eq. \eqref{eq:site_kernels} in the cases of a sphere, cylinder and parallelepiped centered at the origin.
We introduce the short-hand $\mathbf{Q} = \mathbf{G} - \mathbf{G}'+\mathbf{q}$ and let $r_{\mathrm{c}}$ be a sublattice dependent characteristic radius, $h_{\mathrm{c}}$ a cylinder height and let $\mathbf{a}_1,\mathbf{a}_2,\mathbf{a}_3$ subtend the parallelepiped. Then,
\begin{align}
    & \text{Sphere:} \quad \int_{\mathrm{sphere}}e^{-i\mathbf{Q}\boldsymbol{\cdot}\mathbf{r}} \mathrm{d} \mathbf{r} = \frac{4\pi}{Q^2} \left[-r_c \cos(Q r_{\mathrm{c}}) + \frac{\sin(Q r_{\mathrm{c}})}{Q} \right].  \label{eq:K_spherical}
    \\
    & \text{Cylinder:} \quad \int_{\mathrm{cylinder}} e^{-i\mathbf{Q}\boldsymbol{\cdot}\mathbf{r}} \mathrm{d}\mathbf{r} = \frac{4\pi r_{\mathrm{c}}}{Q_zQ_{\rho}}\sin\left(\frac{Q_z h_{\mathrm{c}}}{2}\right) J_1(Q_{\rho} r_{\mathrm{c}}). \label{eq:K_cylinder}
     \\
    & \text{Parallelepiped:} \quad \int_{\text{parallelepiped}} \ee^{-i\mathbf{Q} \boldsymbol{\cdot} \mathbf{r}} \mathrm{d}\mathbf{r} = \Omega_{\mathrm{parallelepiped}} \: \mathrm{sinc}\left(\frac{\mathbf{Q} \boldsymbol{\cdot} \mathbf{a}_1}{2}\right) \mathrm{sinc}\left(\frac{\mathbf{Q} \boldsymbol{\cdot} \mathbf{a}_2}{2}\right) \mathrm{sinc}\left(\frac{\mathbf{Q} \boldsymbol{\cdot} \mathbf{a}_3}{2}\right). \label{eq:K_parallelepiped}
\end{align}
Here $J_1$ denotes a Bessel function of the first kind, $Q=|\mathbf{Q}|$ and $Q_\rho,Q_z$ are projections onto the $\rho$ and $z$-directions of the cylinder.

\twocolumngrid

\bibliography{references.bib}

\end{document}